\documentclass{vgtc}     
\usepackage{multirow}                     

\ifpdf
  \pdfoutput=1\relax                   
  \usepackage{graphicx}                
  \DeclareGraphicsExtensions{.pdf,.png,.jpg,.jpeg} 
\else
  \usepackage{graphicx}                
  \DeclareGraphicsExtensions{.eps}     
\fi%

\graphicspath{{figures/}{pictures/}{images/}{./}} 

\usepackage{microtype}                 
\PassOptionsToPackage{warn}{textcomp}  
\usepackage{textcomp}                  
\usepackage{mathptmx}                  
\usepackage{times}                     
\usepackage{cite}                      
\usepackage{tabu}                      
\usepackage{booktabs}                  
\newcommand{\orcid}[1]{\href{https://orcid.org/#1}{\includegraphics[width=10px]{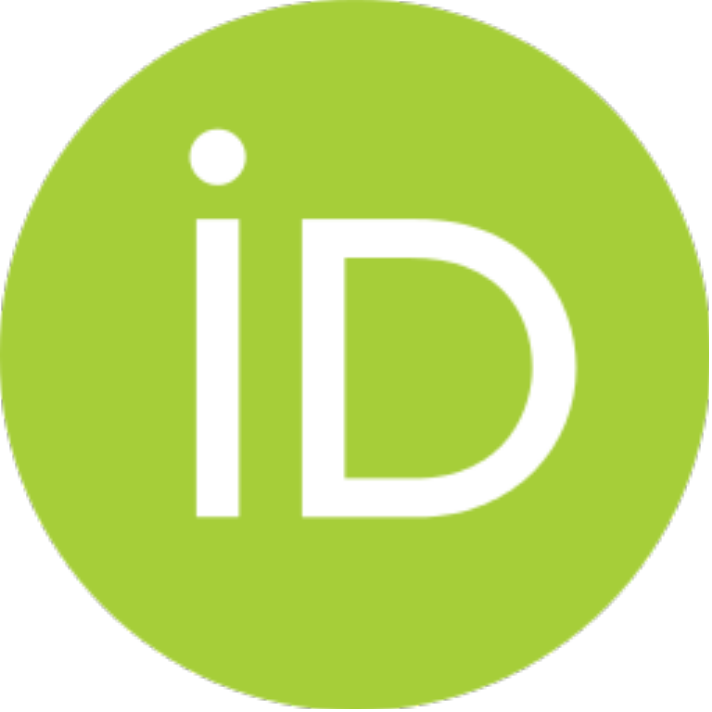}}}
\usepackage{paralist}
\usepackage{tikz}
\usepackage{collcell}
 
\newcommand*{\MinNumber}{0}%
\newcommand*{\MaxNumber}{20}%

 \newcommand{\ApplyGradientK}[1]{%
        \pgfmathsetmacro{\PercentColor}{{100.0*((min(#1,\MaxNumber))-\MinNumber)/(\MaxNumber-\MinNumber)}}
        \colorbox{blue!\PercentColor!white}{\parbox{1em}{\hfill\ifthenelse{#1>10}{\color{white}#1}{#1}}}
}

 \newcommand{\ApplyGradientM}[1]{%
        \pgfmathsetmacro{\PercentColor}{100.0*((min(#1,\MaxNumber))-\MinNumber)/(\MaxNumber-\MinNumber)}
        \colorbox{green!\PercentColor!white}{\parbox{1em}{\hfill#1}}
}

\newcommand{\ApplyGradientN}[1]{%
        \pgfmathsetmacro{\PercentColor}{100.0*((min(#1,\MaxNumber))-\MinNumber)/(\MaxNumber-\MinNumber)}
        \colorbox{red!\PercentColor!white}{\parbox{1em}{\hfill\ifthenelse{#1>10}{\color{white}#1}{#1}}}
}

\newcommand{\ApplyGradientY}[1]{%
        \pgfmathsetmacro{\PercentColor}{100.0*((min(#1,\MaxNumber))-\MinNumber)/(\MaxNumber-\MinNumber)}
        \colorbox{yellow!\PercentColor!white}{\parbox{1em}{\hfill#1}}
}

\newcolumntype{K}{>{\collectcell\ApplyGradientK}r<{\endcollectcell}}
\newcolumntype{M}{>{\collectcell\ApplyGradientM}r<{\endcollectcell}}
\newcolumntype{N}{>{\collectcell\ApplyGradientN}r<{\endcollectcell}}
\newcolumntype{Y}{>{\collectcell\ApplyGradientY}r<{\endcollectcell}}
\renewcommand{\arraystretch}{1}
\usepackage{amssymb,amsmath}
\usepackage{xspace}
\usepackage{paralist}
\usepackage{ifthen}
\usepackage[normalem]{ulem} 
\usepackage{xcolor}
\usepackage[all]{nowidow}

\newboolean{showedits}
\setboolean{showedits}{true} 
\ifthenelse{\boolean{showedits}}
{
  \newcommand{\del}[1]{\textcolor{red}{\sout{#1}}} 
}{
  \newcommand{\del}[1]{} 
  
}
\newboolean{showcomments}
\setboolean{showcomments}{true}

\ifthenelse{\boolean{showcomments}}
{\newcommand{\nbc}[3]{
 {\colorbox{#3}{\bfseries\sffamily\scriptsize\textcolor{white}{#1}}}
 {\textcolor{#3}{\sf\small$\blacktriangleright$\textit{#2}$\blacktriangleleft$}}}
 }
{\newcommand{\nbc}[3]{}
 \renewcommand{\del}[1]{} 
 }

\newcommand{\ie}{\emph{i.e.},\xspace}
\newcommand{\eg}{\emph{e.g.},\xspace}
\newcommand{\etal}{\emph{et al.}\xspace}

\onlineid{0}

\vgtccategory{Research}

\vgtcinsertpkg
\PassOptionsToPackage{pdftex,colorlinks=true,pdfstartview=FitV,
  linkcolor=black,citecolor=black,urlcolor=black,bookmarks=false}{hyperref}

\title{Evaluating Mixed and Augmented Reality: \\A Systematic Literature Review (2009--2019)}

\author{Leonel Merino\thanks{leonel.merino@visus.uni-stuttgart.de; 0000-0002-5396-487X} ~\orcid{0000-0002-5396-487X}\\ %
    \scriptsize University of Stuttgart %
\and Magdalena Schwarzl\thanks{magdalena.schwarzl@visus.uni-stuttgart.de}\\ %
    \scriptsize University of Stuttgart %
\and Matthias Kraus\thanks{matthias.kraus@uni-konstanz.de}\\ %
    \scriptsize University of Konstanz %
\and Michael Sedlmair\thanks{michael.sedlmair@visus.uni-stuttgart.de; 0000-0001-7048-9292} ~\orcid{0000-0001-7048-9292}\\ %
    \scriptsize University of Stuttgart
\and Dieter Schmalstieg\thanks{schmalstieg@tugraz.at; 0000-0003-2813-2235} ~\orcid{0000-0003-2813-2235}\\ %
    \scriptsize Graz University of Technology
\and Daniel Weiskopf\thanks{daniel.weiskopf@visus.uni-stuttgart.de; 0000-0003-1174-1026} ~\orcid{0000-0003-1174-1026}\\ %
    \scriptsize University of Stuttgart}

\teaser{
  \centering
  \includegraphics[width=.95\linewidth]{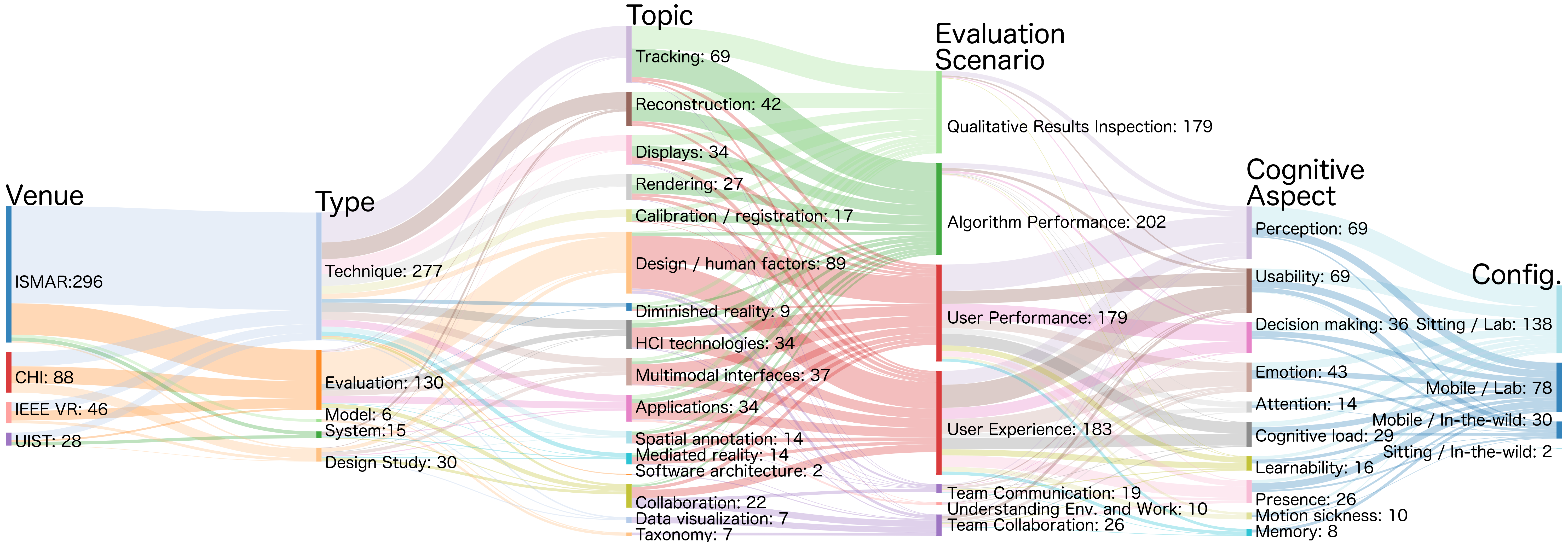}
  \caption{The 458 papers that report on evaluations in mixed and augmented reality (MR/AR) published in ISMAR, CHI, IEEE VR, and UIST between 2009 and 2019 classified by (left-to-right): venue, paper type, research topic, evaluation scenario, cognitive aspects involved in user studies, and study configuration (conducted in the lab or in-the-wild while participants were static or mobile).
 }
  \label{fig:teaser}
}

\abstract{We present a systematic review of 458 papers that report on evaluations in mixed and augmented reality (MR/AR) published in ISMAR, CHI, IEEE VR, and UIST over a span of 11 years (2009--2019). Our goal is to provide guidance 
for future evaluations of MR/AR approaches. To this end, we characterize publications by paper type (\eg technique, design study), research topic (\eg tracking, rendering), evaluation scenario (\eg algorithm performance, user performance), cognitive aspects (\eg perception, emotion), and the context in which evaluations were conducted (\eg lab vs.~in-the-wild). 
We found a strong coupling of types, topics, and scenarios. We observe two groups:
\begin{inparaenum}[(a)]
    \item technology-centric performance evaluations of algorithms that focus on improving tracking, displays, reconstruction, rendering, and calibration, and 
    \item human-centric studies that analyze implications of applications and design, human factors on perception, usability, decision making, emotion, and attention.
\end{inparaenum}
Amongst the 458 papers, we identified 248 user studies that involved 5,761 participants in total, of whom only 1,619 were identified as female. 
We identified 43 data collection methods used to analyze 10 cognitive aspects. 
We found nine objective methods, and eight methods that support qualitative analysis.
A majority (216/248) of user studies are conducted in a laboratory setting. Often (138/248), such studies involve participants in a static way. However, we also found a fair number (30/248) of in-the-wild studies that involve participants in a mobile fashion.
We consider this paper to be relevant to academia and industry alike in presenting the state-of-the-art and guiding the steps to designing, conducting, and analyzing results of evaluations in MR/AR.

} 

\CCScatlist{ 
 \CCScat{I.3.7}{Computing Methodologies}{Computer Graphics}%
{Three-Dimensional Graphics and Realism};
 \CCScat{A.1}{General Literature}{Introductory and Survey}
}

\begin{document}


\firstsection{Introduction}

\maketitle

Across multiple domains, there is an increasing interest in investigating approaches that employ mixed reality (MR) and augmented reality (AR) technologies~\cite{Schmalstieg2016}. 
One example is data visualization, in which researchers of the emerging immersive analytics~\cite{Marr18a} domain study how adopting MR/AR technologies can boost the effectiveness of displaying information and interacting with visualizations. 
However, designing appropriate evaluations that examine MR/AR is challenging, and suitable guidance to design and conduct evaluations of MR/AR are largely missing.
There are several strategies that can be adopted to evaluate MR/AR approaches. When the subject of the evaluation is an algorithm or a novel method, benchmarks can help analyze increases in performance. Sometimes, when a user interface is involved, user studies can provide rich data for the analysis not only of user performance but also of user experience. If the focus of the evaluation is on user environments and work places, surveys and case studies can provide important insights. However, generally it is difficult to identify a suitable evaluation strategy, variables to be examined, and adequate methods to collect relevant data in an evaluation for answering a particular research question. 

Often, approaches are considered most effective when they boost users' performance in decision making. However, we observe that there are several other cognitive aspects (\eg perception, emotion, presence, cognitive load, attention, learnability, and memory) that can also play a fundamental role in the effectiveness of approaches in MR/AR. We conjecture that there is an interplay of cognitive aspects that require evaluations to be comprehensive, for instance, to understand the reasons that led to a high user performance. 
Our goal is to better understand MR/AR evaluation practices with an eye toward guidance on when to perform which type of evaluation.
%

To address our goal we conducted a systematic literature review. We concentrated on the analysis of papers published in ISMAR~\cite{ismar},  CHI~\cite{chi}, IEEE VR~\cite{ieeevr}, and UIST~\cite{uist}. 
We consider these to be the leading venues in MR/AR research, offering a sound and representative body of literature for MR/AR research. We confirm our impressions based on the flagship \emph{A$^*$} classification that ISMAR and CHI and the \emph{A} classification that IEEE VR and UIST obtain in the CORE ranking~\cite{core} (which considers various indicators such as citation rate, paper submissions, and acceptance rate). 
We opted to select papers published in the recent past, and analyzed proceedings of these main MR/AR conferences published across. 

To facilitate the analysis of the papers, we relied on our experience and adopted a popular classification from the visualization community~\cite{ieeevis}, putting papers into one of five types based on their main contribution. 
We observe that MR/AR encompasses multiple topics. Consequently, we complemented the classification by paper type with sixteen research topics that emerged from our analysis. 
 We also adapted the seven evaluation scenarios introduced by Lam~\etal~\cite{Lam12a} (of which we excluded one), and the scenario extended by Isenberg~\etal~\cite{Isen13a}, to the context of MR/AR approaches. In the end, we classified the scenario of evaluations into one of seven types. 
For evaluations that involve users, we identified whether the evaluation was conducted in-the-wild (the targeted real-world usage environment) or in a laboratory, and whether participants of such studies used MR/AR while they were sitting or in a mobile way.
To analyze the implications of approaches that use MR/AR in human cognition, we inferred ten cognitive aspects based on the data collection methods employed in user studies. 
An overview of the relationships of the analyzed dimensions is presented in~\autoref{fig:teaser}.
%
%
%
       The main contribution of our paper is threefold:
        \begin{inparaenum}[(a)]
            \item a systematic analysis of paper types, research topics, evaluation scenarios, cognitive aspects that emerge from data collection methods, and configurations of evaluations in MR/AR,
            \item a synthesis that describes implications for evaluating MR/AR approaches, and
            \item a publicly available data set of the data collected in our systematic literature analysis~\cite{dataset}.
        \end{inparaenum}


\section{Related Work}
\label{sec:related}
To elaborate on the related work, we discuss previous papers that cover various aspects of evaluations in MR/AR. Next, we leverage our experience in visualization research 
and extend our analysis to visualization studies that share our focus on evaluations. Finally, we elaborate on commonalities and differences of these related works to our investigation.   

There are a number of survey articles in MR/AR. Swan and Gabbard~\cite{swan2005survey} surveyed AR papers published in 1992--2004 and analyzed 21 papers that describe user evaluations. Similarly, a few years after, Duenser~\etal~\cite{Duen08a} analyzed 161 AR papers published in 1993--2007. Both studies classified papers by evaluation type (\eg perception, performance) and involved methods (\eg objective/subjective measurements, qualitative analysis). They found that 47\% of user evaluations measured user task, and 22\% analyzed variables of perception or cognition. We consider these works complementary to our study. 
Zhou~\etal~\cite{Zhou08a} focused their literature review on tracking, interaction, and display technologies. They found an emerging trend of papers that focused primarily on evaluations, which accounted to 5.8\% of the reviewed papers. 
Kruijff~\etal~\cite{5643530} presented a classification of perceptual issues grouped into categories such as environment, capturing, augmentation, display, and individual user differences. 
Fite-Georgel~\cite{6162889} surveyed AR industrial applications, organized into categories that relate to the stages of the life-cycle of products. The applications were evaluated using the following criteria:
    workflow integration,
    scalability,
    cost/benefit,
    out of the lab,
    user tested,
    out of developers' hands, and
    involvement of the industry.
Radu and MacIntyre~\cite{6402561} investigated how AR designs relate to children's skills, such as motor abilities, spatial cognition, attention, logic, and memory. 
Krichenbauer~\etal~\cite{6948405} surveyed professionals who create 3D media content using AR user interfaces. A set of requirements were distilled and implemented in a prototype tool. 
Grubert~\etal~\cite{7435333} presented a taxonomy for pervasive AR and context-aware AR based on context sources, context targets, and context controllers. 
Chen~\etal~\cite{8115411} classified medical MR to identify areas with little research as well as to provide references to practitioners. 
Recently, Kim~\etal~\cite{Kim18a} reviewed the literature in MR/AR published in 2008--2017, and found a sharp increase in AR evaluation to which they related 16.4\% of the reviewed papers. 
Fonnet and Prié~\cite{8770302} surveyed 177 immersive analytics papers published in 1991--2018. They included in the analysis aspects of evaluations such as immersion, technologies, interaction, and visualization techniques. 
Dey \etal~\cite{Dey18a} reviewed the MR/AR research literature that reports on user studies published in 2004--2014. They found an increasing trend of involving handhelds in AR user studies. They also confirmed that most user studies are conducted in laboratory settings. In contrast, our study includes more recent papers and elaborates on a broader view that includes both human- and technology-based evaluations.  

There are studies in other fields that reviewed their respective literature and analyzed evaluations. We name a few examples from the field of visualization: 
Carpendale~\cite{Carp08a} discussed characteristics of information visualization evaluation in terms of evaluation strategies, data collection methods, and analysis methods. She reflected on the need for conducting more evaluations and postulated that evaluation should be more diverse in terms of employed methodologies.     
Lam~\etal~\cite{Lam12a} identified seven scenarios of information visualization evaluation. The scenarios encapsulate current evaluation practices, which can guide researchers to design more effective evaluations. 
Isenberg~\etal~\cite{Isen13a} later expanded the scenarios to include evaluations based on qualitative results inspection. 
Elmqvist and Yi~\cite{Elmq15a} proposed a set of general and reusable patterns to commonly occurring problems in evaluating visualization approaches.
Merino~\etal~\cite{Meri18a} found that 62\% of software visualization evaluations involved  usage examples and anecdotal evidence,  29\% experiments, and 7\% case studies. Our work is methodically inspired by these systematic analyses,
which we apply to MR/AR for the first time. 
In particular, our coding scheme uses the paper types described by Munzner~\cite{Munz08a} and seven of the evaluation scenarios defined by Lam~\etal~\cite{Lam12a} and Isenberg~\etal~\cite{Isen13a}.
\section{Methodology}
\label{sec:method}

We employed a systematic literature review approach
.  To mitigate potential biases in the results of the survey, we followed the comprehensive guidelines by Kitchenham~\cite{Kitc02a}. 
The methodology offers robust and transferable evidence for evaluating and interpreting relevant research on a topic of interest. To this end, we defined a review protocol to ensure rigor and reproducibility, in which we determine
\begin{inparaenum}[(a)]
    \item a data collection method, 
    \item selection criteria, 
    \item a coding scheme, and 
    \item a coding process.
\end{inparaenum}

\subsection{Data Collection Method}
We collected papers published in ISMAR, CHI, IEEE VR, and UIST.
To find primary studies for our analysis, we collected all papers in the proceedings of ISMAR of the period 2009--2019. 
Next, we used the ACM Digital Library~\cite{acmdl} to collect papers from CHI. We used IEEE Xplore~\cite{ieeeexplor} to collect papers from IEEE VR and UIST. In neither case, we included keywords such as ``evaluation'' in the search. That is, we first identified MR/AR papers, and then manually analyzed evaluations. 

\subsection{Selection Criteria}
\begin{figure}[t]
  \centering
  \includegraphics[width=\linewidth]{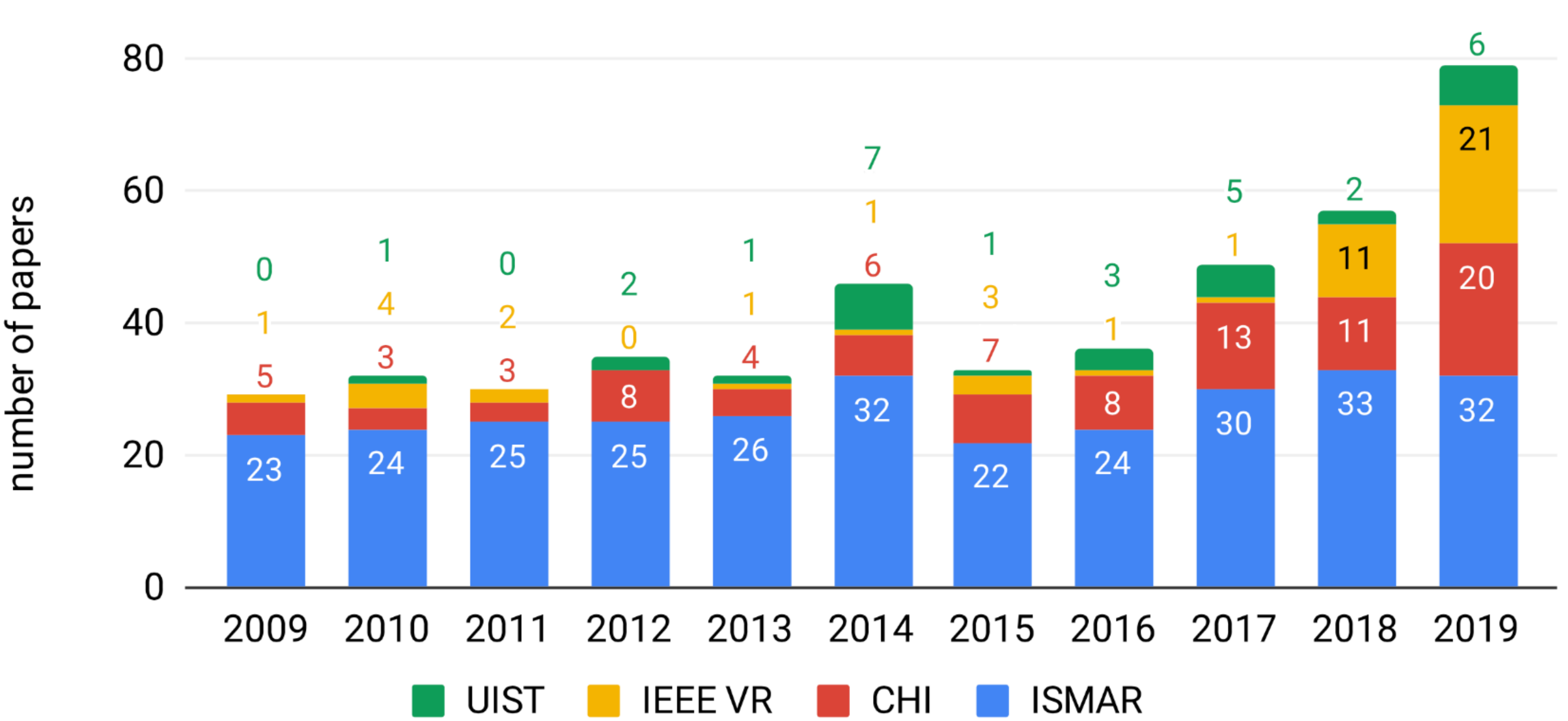}
  \caption{The 458 included papers by publication year and venue.}
  \label{fig:papers}
\end{figure}

We analyzed the proceedings of 11 years (2009--2019) of ISMAR and included 296 papers. These papers correspond to full and short papers from 2009 until 2014, and T\&S conference and TVCG journal papers from 2015 until 2019. We excluded other publication formats that, due to their brevity, are unlikely to contain enough details regarding an evaluation (\eg posters, demos, keynotes, extended abstracts).
Also, we collected 88 papers from CHI, 46 papers from IEEE VR, and 28 papers from UIST. Since these venues not only focus on MR/AR, we excluded papers that either focus on a different topic (\eg virtual reality) or do not report on evaluation explicitly. Our set has 458 papers with 4 to 14 pages in length. A temporal histogram of the selected papers is shown in~\autoref{fig:papers}.

\subsection{Coding Scheme}
\label{sec:coding-scheme}
To analyze the evaluations reported in the MR/AR literature, we coded paper types, research topics, and evaluation scenarios. In evaluations where we identified users studies, we also coded data collection methods to infer cognitive aspects, number and gender of participants, and the adopted configuration of mobility. 

\vspace{0.2em}
\noindent\textbf{Paper Types.}
We classified the paper type according to categories proposed by Munzner~\cite{Munz08a}. Although these categories aim at characterizing visualization papers, we observed that the categories can be generalized to MR/AR papers. As this classification has been widely adopted in multiple studies in visualization research, using it in our study of MR/AR evaluations provides a bridge between the two fields that enables comparison.
In summary, a paper can be classified into one of five types from~\autoref{tab:types}.

\begin{table}[!ht]
\caption{Paper Types.}
\label{tab:types}
\setlength\tabcolsep{1pt}
\begin{tabular}{p{1.7cm}p{6.7cm}} \toprule
Technique   & Papers focusing on new algorithms that improve the performance of an approach. \\[2pt]
Evaluation  & Papers that elaborate on a judgment of the quality, importance, or value of an approach. They can describe careful examinations of a real-world case (\ie case study) or the behavior of users exposed to a tool (\ie user study).\\[2pt]
System     & Papers that elaborate on choices made in the design of the architecture of a proposed system or framework, and lessons learned from its use. These can be seen as meta-techniques that enable the generation of new techniques. \\ [2pt]
Model   & Papers that include 
\begin{inparaenum}[(a)]
    \item commentaries of an expert in the field who argues to support a position,
    \item formalisms of models, definitions, or terminology related to techniques, and 
    \item taxonomies and categories to help researchers analyze a domain.
\end{inparaenum}\\[2pt]
Design study/ Application & Papers that describe how existing techniques can be useful to deal with a concrete problem in a domain.\\
\bottomrule                           
\end{tabular}
\end{table}



\vspace{0.2em}
\noindent\textbf{Research Topics.}
There are several research topics relevant to MR/AR. 
Although there are previous classifications of topics~\cite{Zhou08a,Kim18a}, 
we opted to identify topics by ourselves. We think that comparing our resulting list to previous ones can help confirm the relevance of topics in common and identify as emergent topics the categories that are different. To this end, we analyzed topics listed in calls for papers and complemented the list with topics that emerge from the analysis of paper titles and keywords. 
For each paper included in our study, we identified one main research topic.
In the end, we defined the 16 topics listed in~\autoref{tab:topics}. 

\vspace{0.2em}
\noindent\textbf{Evaluation Scenarios.}
For each paper in our analysis, we looked for details of an evaluation, and, when we found some details, we classified the evaluation \emph{scenario}. A scenario is the context in which an evaluation is carried out. 
We originally considered the eight scenarios that characterize the context in which evaluations are conducted in visualization~\cite{Lam12a,Isen13a}. We observed that, while most scenarios fit well the context of MR/AR, ``Visual Data Analysis and Reasoning'' is too specific to the visualization domain, and therefore we excluded it. We adapted the remaining seven scenarios to the context of evaluation in MR/AR. 
We furthermore grouped these scenarios according to the 
level of user involvement in an evaluation:
\begin{inparaenum}[(a)] 
    \item \emph{technique-centered} scenarios do not involve users,
    \item \emph{user-centered} scenarios involve users who individually interact with a technique, and 
    \item \emph{team-centered} scenarios involve users who  interact with each other with the support of a technique (often simultaneously).
\end{inparaenum}  
\autoref{tab:scenarios} presents the scenarios adapted to the evaluation of MR/AR approaches.

\begin{table}[t]
\caption{Evaluation Scenarios.}
\label{tab:scenarios}
\renewcommand\arraystretch{0}
\begin{tabular}{p{0.4cm}p{0.8cm}p{7.6cm}} \toprule
\parbox[t]{2mm}{\multirow{2}{*}[-2ex]{\rotatebox[origin=c]{90}{Technology}}} & AP &  Algorithm Performance: a quantitative evaluation of the technical performance, typically using benchmarks to compare rendering speed or memory performance.\\[2pt]
                           & QRI & Qualitative Results Inspection: a qualitative discussion of results that encourages readers to agree on a quality statement. \\[3pt]
                          
\parbox[t]{2mm}{\multirow{2}{*}[-5ex]{\rotatebox[origin=c]{90}{Human}}}      & UP & User Performance: a quantitative or qualitative evaluation of the performance of the users of a system. Typically, user performance is measured in experiments using time and correctness of users to complete as set of tasks. \\[2pt]
                           & UE & User Experience: an examination of how a user reacts to interacting with a tool. Likert scale questionnaires for subjective feedback, and interviews are commonly used to gather data of user experience. \\[3pt] 
\parbox[t]{2mm}{\multirow{3}{*}[-9ex]{\rotatebox[origin=c]{90}{Team}}}      & UWP & Understanding environment and Work Practices: an examination to understand the implications of adopting a technique into a working environment. Common examples of data collection methods are surveys and interviews with expert users. \\[2pt]
                           & COM & Team communication in MR/AR: an assessment of the communicative value of a technique in regards to goals such as teaching or presentation. \\[2pt]
                           & COL & Team collaboration in MR/AR: an evaluation of the level of support of a technique to facilitate collaboration in a team.\\ 

\bottomrule                           
\end{tabular}
\end{table}

\begin{table}[!ht]
\caption{Research Topics.}
\label{tab:topics}
\setlength\tabcolsep{1pt}
\begin{tabular}{p{1.6cm}p{6.7cm}} \toprule
Tracking&Papers evolving around 3D tracking. It 
also contains most papers dealing with simultaneous localization and 
mapping, if the emphasis is on localization.\\[2pt]
Reconstruction&Technical papers focusing on 3D reconstruction, 
either as a prerequisite for MR/AR applications (which will typically 
use the reconstructed models to derive some form of spatial annotation), 
or SLAM papers where the mapping part is most relevant.\\[2pt]
\multicolumn{1}{l}{\begin{tabular}[t]{@{}l@{}}Calibration/\\registration\end{tabular}}&Papers focusing on spatial registration for real-time tracking. These papers have a thematic overlap with tracking and reconstruction.\\[2pt]
Rendering&Papers dealing with coherent rendering, in 
particular, global illumination for MR, inverse rendering, and photometric registration.\\[2pt]
Displays&Papers that deal with physical displays for MR/AR, mostly head-mounted displays and spatial AR.\\[2pt]
HCI~technologies&Papers discussing technical solutions to interaction problems.\\[2pt]
\multicolumn{1}{l}{\begin{tabular}[t]{@{}l@{}}Design/\\human factors\end{tabular}}&Papers dealing with the design (and  evaluation) of interaction techniques or with the study of human factors per se that occur in the context of MR/AR systems. One important group are perceptual issues, in particular, depth perception.\\[2pt]
Applications&Papers exploring MR/AR interfaces in specific application use cases, covering both medical and non-medical applications.\\[2pt]
Multimodal interfaces&Papers dealing with audio, haptics, and other non-visual modalities.\\[2pt]
Collaboration&Papers describing collaborative MR/AR.\\[2pt]
\multicolumn{1}{l}{\begin{tabular}[t]{@{}l@{}}Mediated\\reality\end{tabular}}&Papers on changing the appearance of physical objects and scenes.\\[2pt]
\multicolumn{1}{l}{\begin{tabular}[t]{@{}l@{}}Spatial\\annotation\end{tabular}}&Papers that display semantic information registered to the real world, to instruct or guide the user. The main difference to mediated reality is that the real objects remain mostly 
visible and are ``augmented'', not ``supplanted''.\\[2pt]
\multicolumn{1}{l}{\begin{tabular}[t]{@{}l@{}}Data\\visualization\end{tabular}}&Papers that elaborate on the display of data registered to the real world in an MR/AR display. Difference to spatial annotation is that the data undergoes a noteworthy visual encoding, as opposed to annotations, which are visually trivial in most cases (such as a colored icon or text label).\\[2pt]
Diminished reality&Papers on all kinds of techniques that make real things disappear or partially transparent.\\[2pt]
Taxonomy&Papers describing theoretical discussions and taxonomies.\\[2pt]
\multicolumn{1}{l}{\begin{tabular}[t]{@{}l@{}}Software\\architecture\end{tabular}}&Papers describing software architectures.\\
\bottomrule
\end{tabular}
\end{table}

\vspace{0.2em}
\noindent\textbf{Cognitive Aspects.}
There are various aspects of human cognition that can be considered in user evaluations of MR/AR, which can allow researchers to obtain a more comprehensive understanding of the impact of their approaches. 
Commonly, studies focus only on a few cognitive aspects, so the scope of their analyses stays feasible. However, understanding multiple cognitive aspects together can be used to build theories that explain complex phenomena of human factors in MR/AR.
We did not find cognitive aspects explicitly described in evaluations. Therefore, we adopted a bottom-up approach and inferred them from employed data collection methods. We did not collect general data collection methods such as questionnaires or interviews. Instead, we collected methods that are used in evaluations to analyze specific human aspects that deal with cognition. That is, these aspects deal with ``the mental action or process of acquiring knowledge and understanding through thought, experience, and the senses''~\cite{lexico}. Therefore, we used these methods as a proxy to identify aspects of human cognition involved in evaluations. For example, in evaluations that describe the use of the Self-Assessment Manikin~\cite{Brad94a} method, we can infer that researchers investigate \emph{emotions}. In the end, we identified 10 cognitive aspects, presented in~\autoref{tab:cognition}.

\begin{table}[!ht]
\caption{Cognitive Aspects.}
\label{tab:cognition}
\setlength\tabcolsep{1pt}
\renewcommand\arraystretch{0}
\begin{tabular}{p{1.5cm}p{6.7cm}} \toprule
Perception &  Relates to the interpretation of sensory information (\eg visual, auditory, or haptic) to understand information of the environment.\\[3pt]
Usability &  The ease of use. \\[3pt]
Emotion &  A mental state that relates to thoughts, feelings, behavior, and affects.\\[3pt]
Decision making & The process of identifying and choosing from several alternative possibilities. \\[3pt]
Presence &  The feeling of having no mediation between oneself and the (virtual) environment, which promotes the psychological sensation of ``being there''.\\[3pt]
Cognitive load &  Relates to the mental load imposed by instructional parameters, \eg task structure, the sequence of information given during an evaluation; and mental effort that refers to the capacity allocated by participants of a study to the instructional demands.\\[3pt]
Attention &  The process of selectively concentrating on an aspect while ignoring other information.\\[3pt]
Learnability & Capability of a system to enable users to learn how to use it, usually considered as an aspect of usability.\\[3pt]
Motion sickness & A disturbance of the senses due to a difference between actual and expected motion.\\[3pt]
Memory &  Relates to the ability of encoding, storing, and retrieving information when needed.\\
\bottomrule                           
\end{tabular}
\end{table}

\vspace{0.2em}
\noindent\textbf{Study Configurations.}
For each user evaluation, we extract the number and gender of involved participants. 
As MR/AR devices often allow mobile use, we analyze whether this characteristic is present in user evaluations, or whether evaluations are conducted with users in a static way. 
Moreover, we code whether user evaluations are conducted in a laboratory setting or whether they correspond to field studies conducted in-the-wild. In-the-wild studies, a term commonly used in human-computer interaction (HCI), are conducted in a real-life scenario targeted by a MR/AR approach. Notice that in-the-wild does not necessarily imply outdoor usage, as multiple MR/AR approaches target indoor activities. 

\subsection{Coding Process}
The coding process was carried out by the first three co-authors of this paper. Each of them analyzed a similar number of papers. Each paper was reviewed at least by two coders. Coders trained themselves by classifying the paper types of 72 publications of ISMAR in 2014--2018 and reached a ``substantial''~\cite{landis1977measurement} 0.7353 Krippendorff's alpha~\cite{krippendorff2018content} intercoder reliability. For all papers, we crosschecked the results and discussed conflicting results to reach a consensus. 
We built on our experience on visualization research to code paper types and evaluation scenarios using a defined set of categories. Research topics, cognitive aspects, and study configurations emerged from the analysis of papers and were iteratively refined. Categories of paper types, research topics, and study configuration are mutually exclusive, whereas multiple evaluation scenarios and cognitive aspects could be associated to individual papers.      

To classify the papers, we followed an incremental reading approach. We started with the title, keywords, abstracts, and skimming figures, which in many cases already clarified paper types and main research topic. If unclear, we continued reading the introduction, and in some cases the entire paper. For coding evaluation scenario, cognitive aspects, and study configuration, we additionally identified the respective evaluation sections in the paper and closely read those.


\section{Results}
\label{sec:results}


We now report on the results coding the 458 identified MR/AR papers. The results are organized according to the coding scheme introduced in~\autoref{sec:coding-scheme} and~\autoref{fig:teaser}. 
A summary of the results is presented in~\autoref{tab:research_topics} with the number of papers in each research topic classified by paper types, evaluation scenarios, and cognitive aspects. 


\subsection{Paper Types}
\begin{figure}[t]
  \centering
  \includegraphics[width=\linewidth]{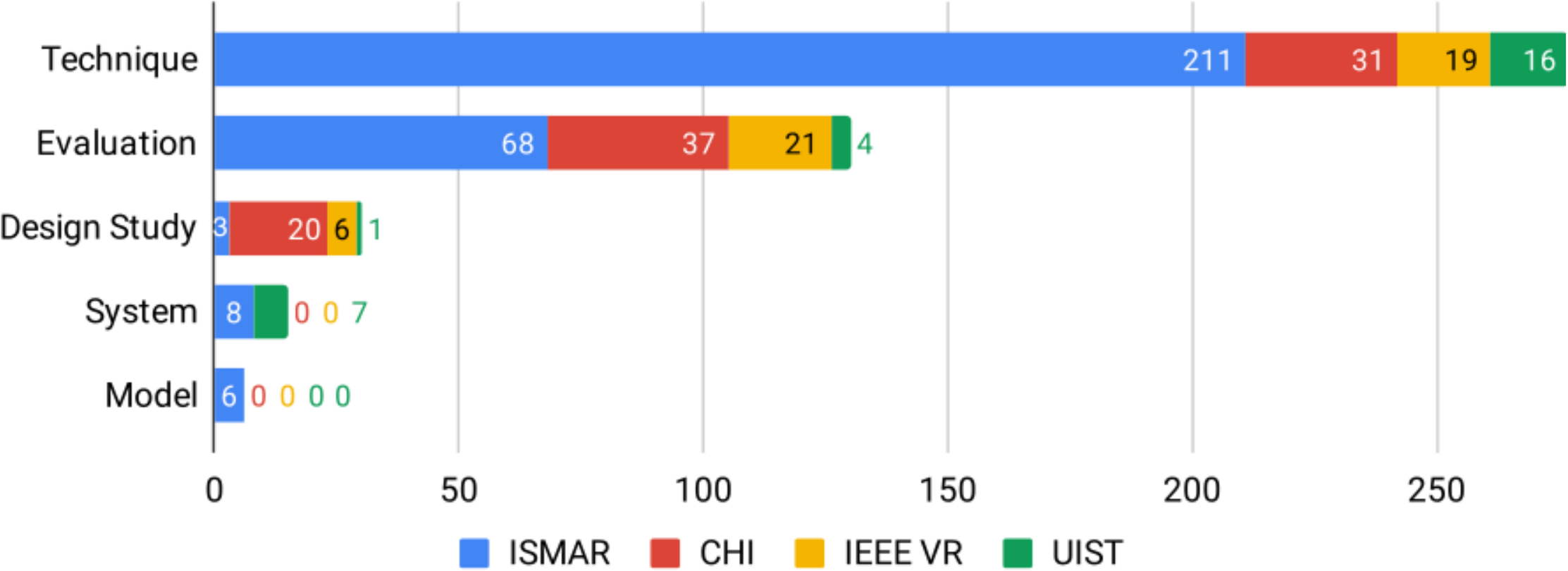}
  \caption{A classification of the 458 papers by type and publication venue: technique, evaluation, design study, system, and model.}
  \label{fig:types}
\end{figure} 
\begin{figure}[t]
  \centering
  \includegraphics[width=\linewidth]{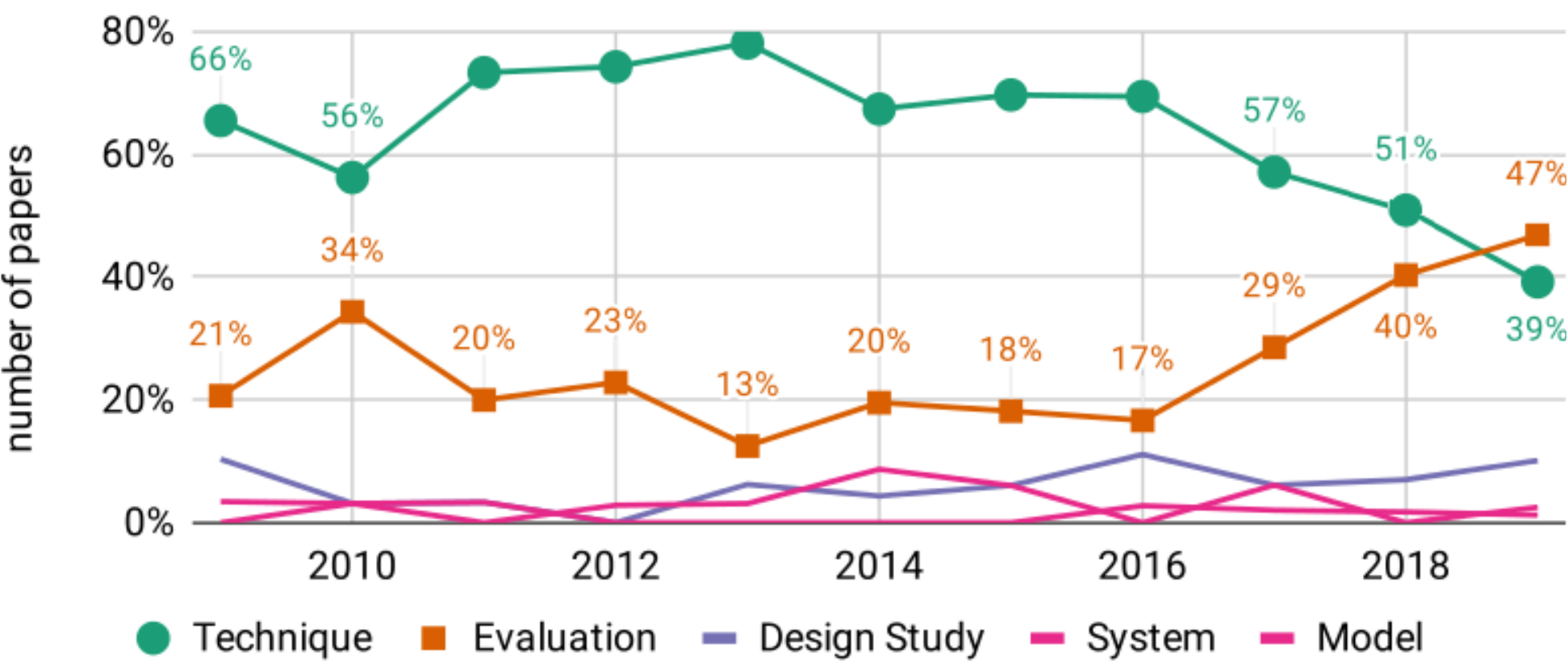}
  \caption{Trends of the types of papers per year.}
  \label{fig:types_evolution}
\end{figure} 
A summary of the results of our classification by paper type is presented in~\autoref{fig:types}. 
We observe that the types of papers vary across venues. In ISMAR papers (296), technique papers (211) outnumber evaluation papers (68) by a factor of three. In CHI papers (88), evaluation papers (37), technique papers (31), and to a lesser extent design study papers (20) are almost balanced. In IEEE VR (46), papers are mostly of two types: evaluation (21) and technique (19). UIST papers (28), mostly consist of technique papers (16) and system papers (7).

We present the percentage of paper types over the total number of published papers during 2009--2019 in~\autoref{fig:types_evolution}. 
Only technique and evaluation papers have non-marginal frequencies. Since there are small differences in the percentage of papers over time, the sum of the percentages of technique and evaluation papers is frequently close to 100\%, making them the core pillars of the MR/AR literature.
We observe that the percentage of evaluation papers is stable but low until 2016, in which a noticeable steady increase appears. In 2019, for the first time the percentage of evaluation papers exceeds the percentage of technique papers. We observe that this increase is triggered by the increased number of evaluation papers published in IEEE VR. 
We think that the increased focus on human-centric papers is a positive symptom of MR/AR becoming a more mature field, in which robust techniques and hardware are increasingly more available.
\subsection{Research Topics}
\begin{figure}[t]
  \centering
  \includegraphics[width=\linewidth]{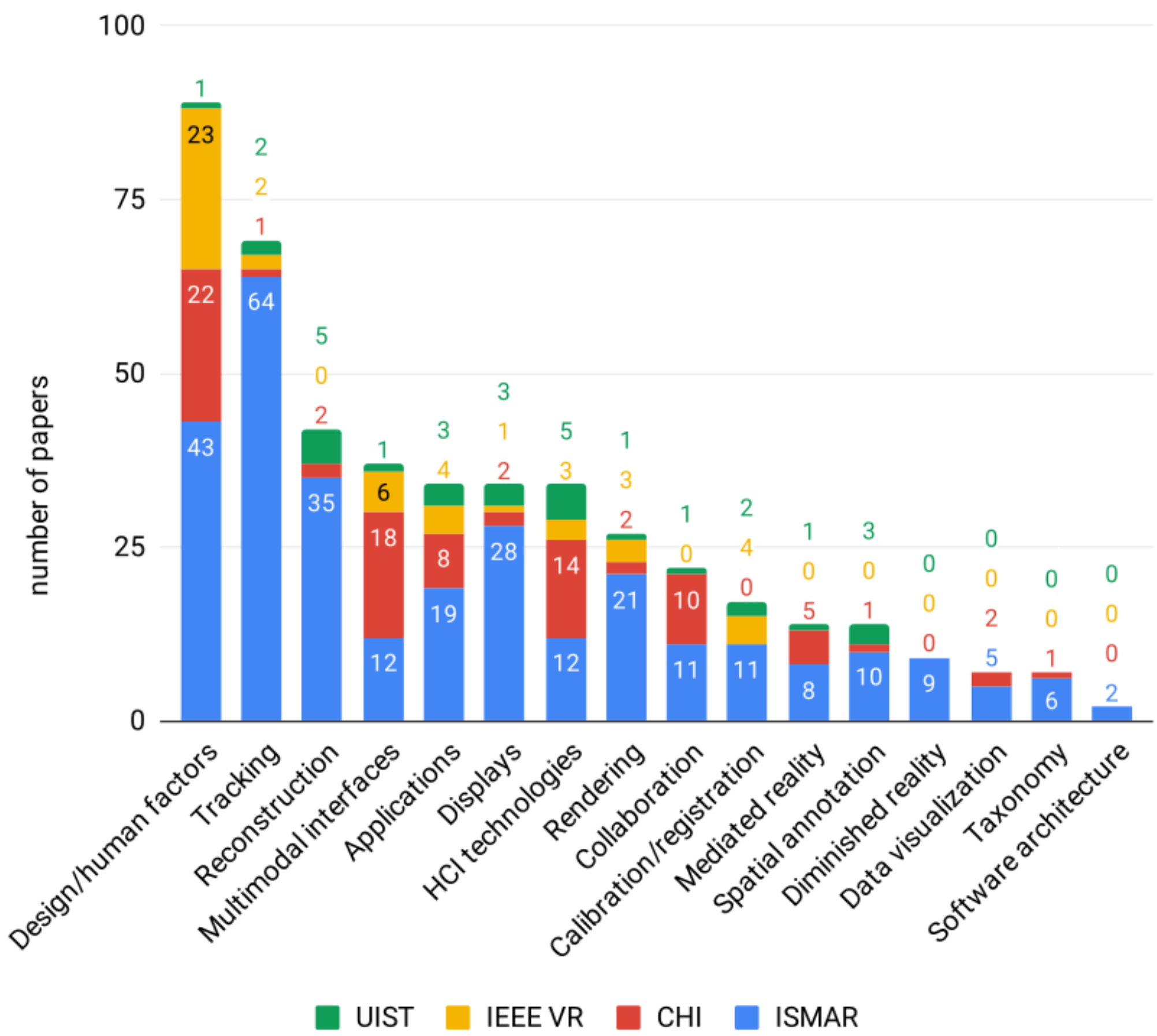}
  \caption{Research topics of interest to MR/AR.}
  \label{fig:topics_percentage}
\end{figure}

We identified 16 different research topics, as summarized in~\autoref{fig:topics_percentage}. 
When comparing the results to a previous study~\cite{Kim18a}, we identified some new topics: design/human factors, mediated reality, spatial annotation, diminished reality, taxonomy, and software architecture. Although there might be an overlap of topics that could explain many differences, some of them could identify emergent topics in MR/AR. The frequency of topics varies amongst venues. In ISMAR papers (296), the main topics are tracking (64), design/human factors (43), reconstruction (35), displays (28), and rendering (21). 
These topics are coherent with the proportion of technique versus evaluation papers that we found. In contrast, the main topics of interest in CHI papers (88) are design/human factors (22), multimodal interfaces (18), HCI technologies (14), collaboration (10), and applications (8), 
which are in line with the balanced number of technique, evaluation, and design study papers. In IEEE VR papers (46), main topics are design/human factors (23), multimodal interfaces (6), applications (4), calibration/registration (4), and rendering (3). In UIST papers (28), the main topics are reconstruction (5), HCI technologies (5), applications (3), displays (3), and spatial annotation (3). IEEE VR and UIST seem to blend the topics of interest of ISMAR and CHI.         
\begin{figure}[t]
  \centering
  \includegraphics[width=\linewidth]{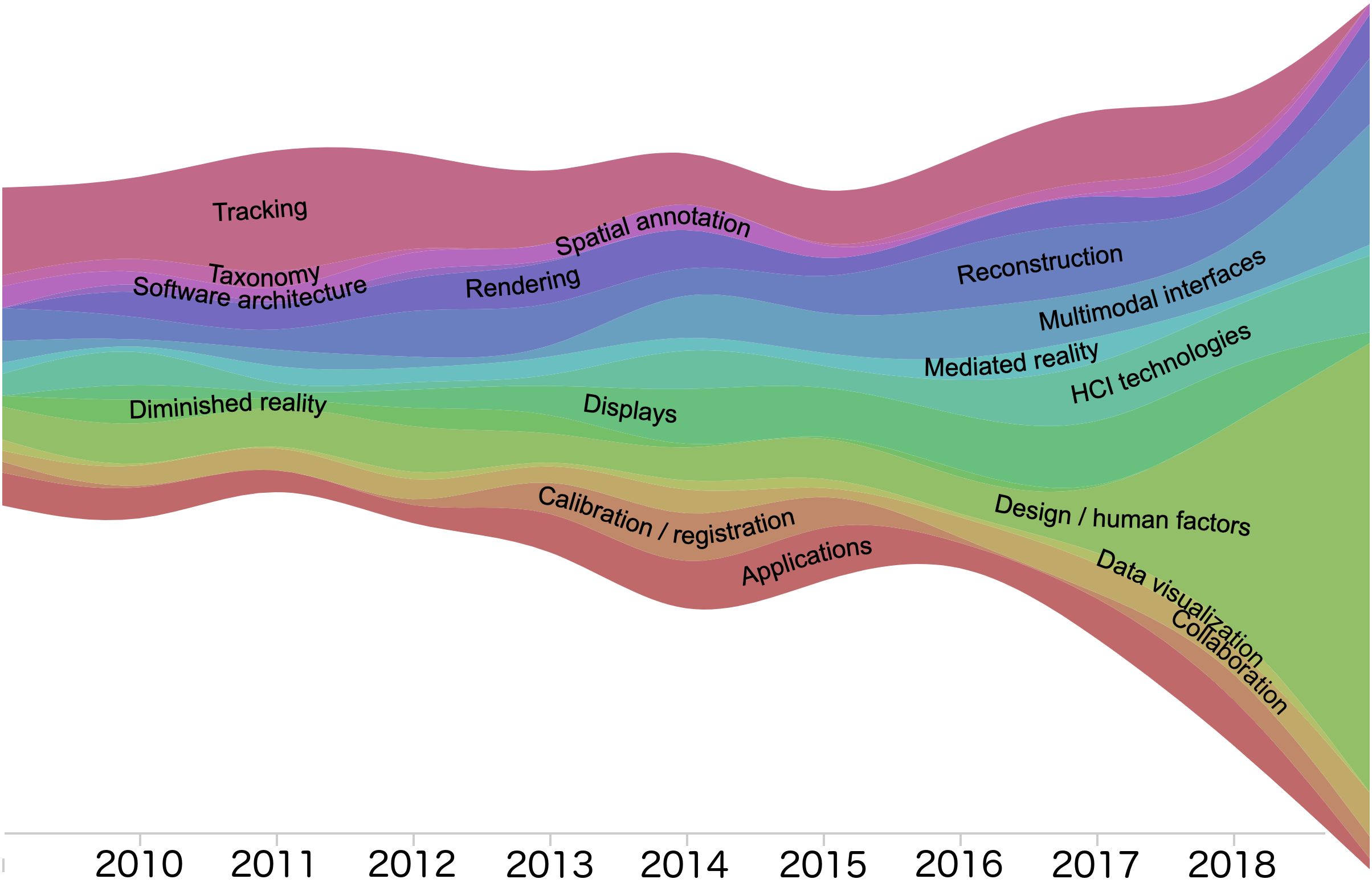}
  \caption{Trends of research topics per year.}
  \label{fig:topics_evolution}
\end{figure} 

\autoref{fig:topics_evolution} shows a chart with the trends of the number of papers by research topic over time that help us analyze emergent topics. Indeed, since 2016, the number of papers dedicated to design/human factors have been greatly increasing and become predominant, exceeding the number of papers that focus on tracking, which exhibit a fairly decreasing trend. In turn, a fair number of papers dedicated to spatial annotation are found between 2009 and 2012, but completely absent after 2015; however, this topic of interest reappeared in 2018--2019. Other topics, such as diminished reality, software architecture, and taxonomy, are intermittent. A steady number of  applications, mediated reality, collaboration, rendering papers are found in MR/AR.

\subsection{Evaluation Scenarios}
\begin{figure}[t]
  \centering
  \includegraphics[width=\linewidth]{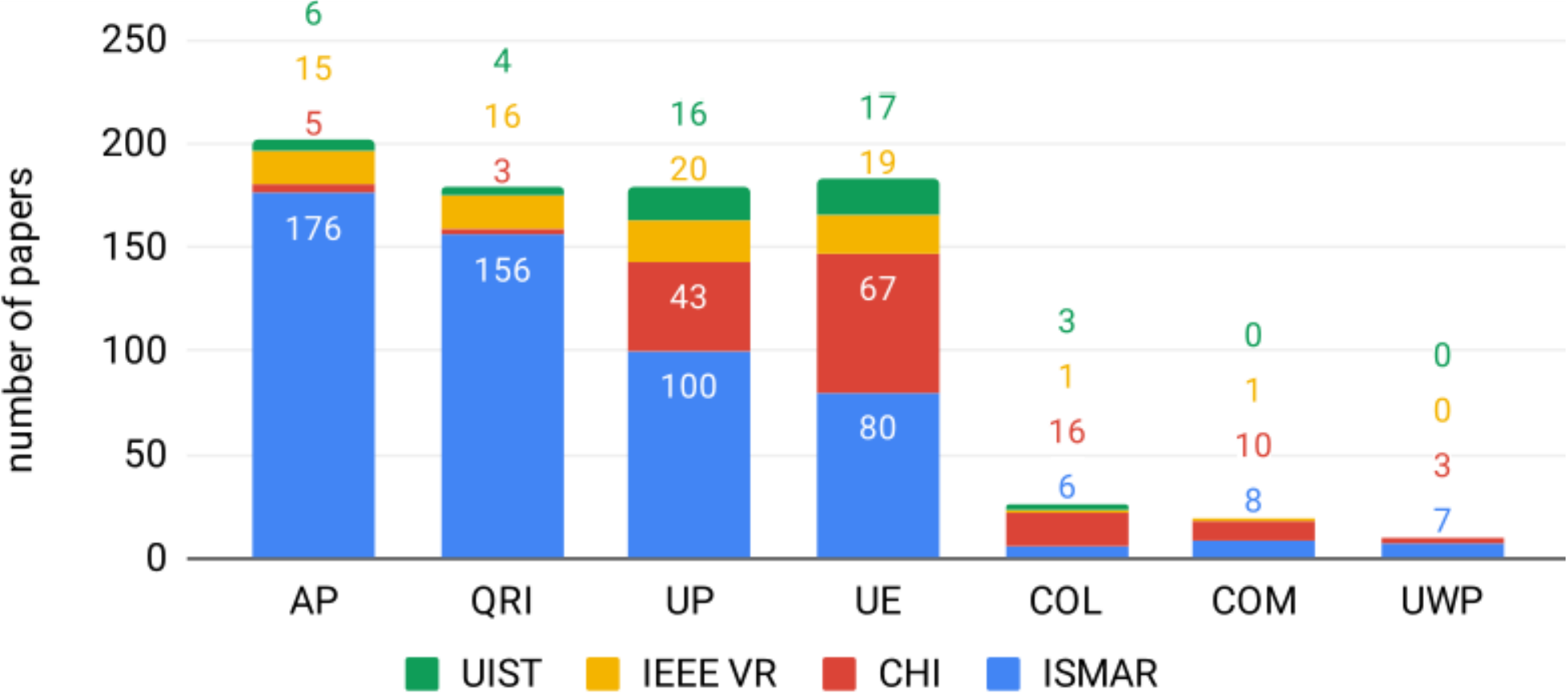}
  \caption{The 801 evaluation scenarios identified among the 458 analyzed papers: Algorithm Performance (AP), Qualitative Results Inspection (QRI), User Performance (UP), User Experience (UE), Understanding environment and Work Practices (UWP), Team Communication (COM), and Team Collaboration (COL).}
  \label{fig:scenarios}
\end{figure} 
\begin{figure}[t]
  \centering
  \includegraphics[width=\linewidth]{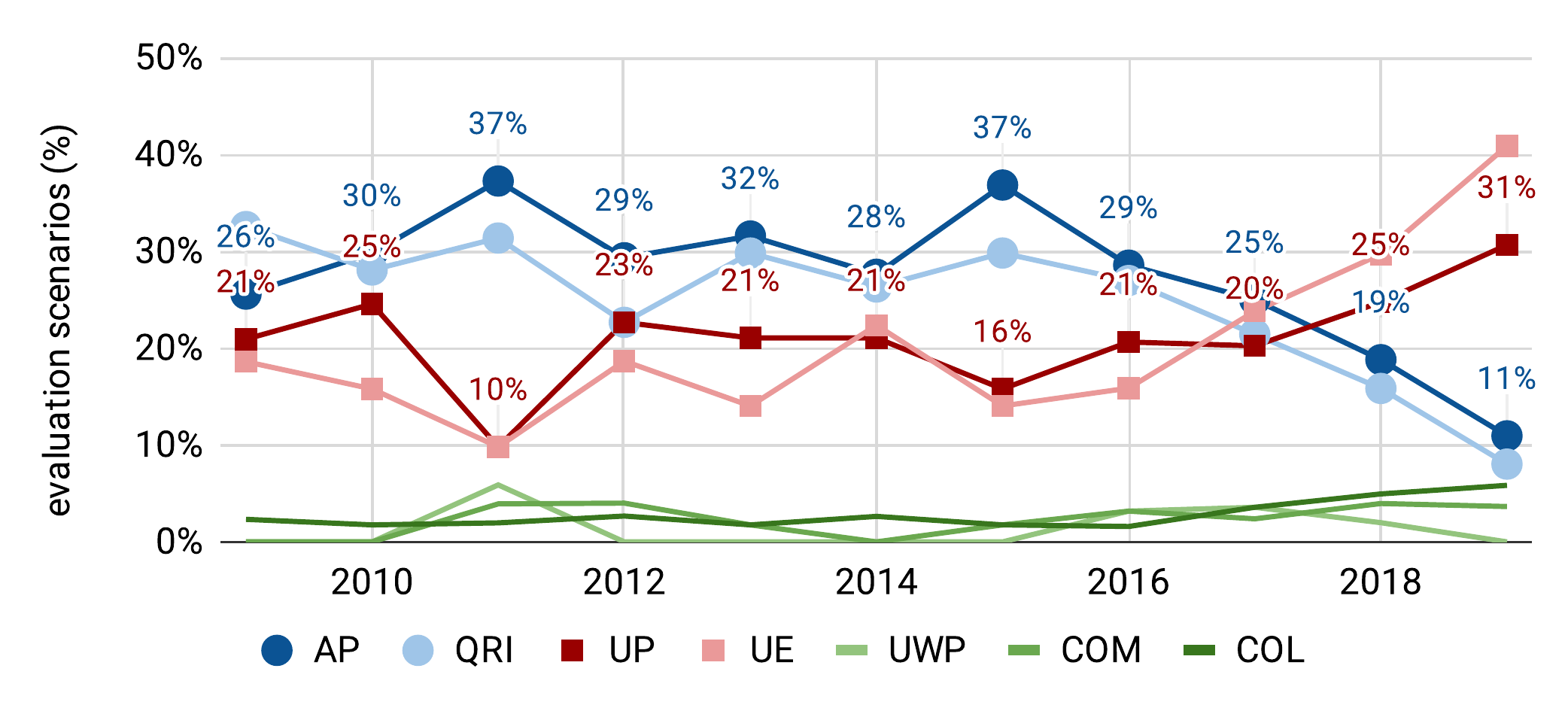}
  \caption{Trends of MR/AR evaluation scenarios (\emph{notice that an evaluation can involve multiple scenarios}).}
  \label{fig:scenarios_evolution}
\end{figure}
We summarize the number of papers per evaluation scenario in~\autoref{fig:scenarios}. As the evaluations in a paper can involve multiple scenarios, we found 801 evaluation scenarios in total, and we confirmed that:
\begin{inparaenum}[(a)]
    \item most papers (382) involve evaluations of technique-centered scenarios (\eg benchmarks),
    \item many papers (363) describe evaluations of user-centered scenarios (\eg user studies), and
    \item a few papers (56) elaborate on evaluations of team-centered scenarios (\eg surveys).
\end{inparaenum}
We observe that evaluations in ISMAR frequently focus on validating techniques involving AP+QRI scenarios (176/296) and design/human factors involving UP+UE scenarios (100/296), and much less frequently involving COL+COM+UWP (8/296). In contrast, evaluations in CHI and UIST mostly involve UP+UE scenarios (84/116), less frequently involve COL+COM+UWP (19/116), and rarely involve AP+QRI (11/116). 


 \autoref{fig:scenarios_evolution} 
shows a line chart with the trends of the seven evaluation scenarios: technique-centered scenarios (AP+QRI) in blue tones at the top, user-centered scenarios (UP+UE) in red tones in the middle, and team-centered scenarios (UWP+COM+COL) in green tones at the bottom of the chart.
\subsection{Cognitive Aspects}
\begin{figure}[t]
  \centering
  \includegraphics[width=\linewidth]{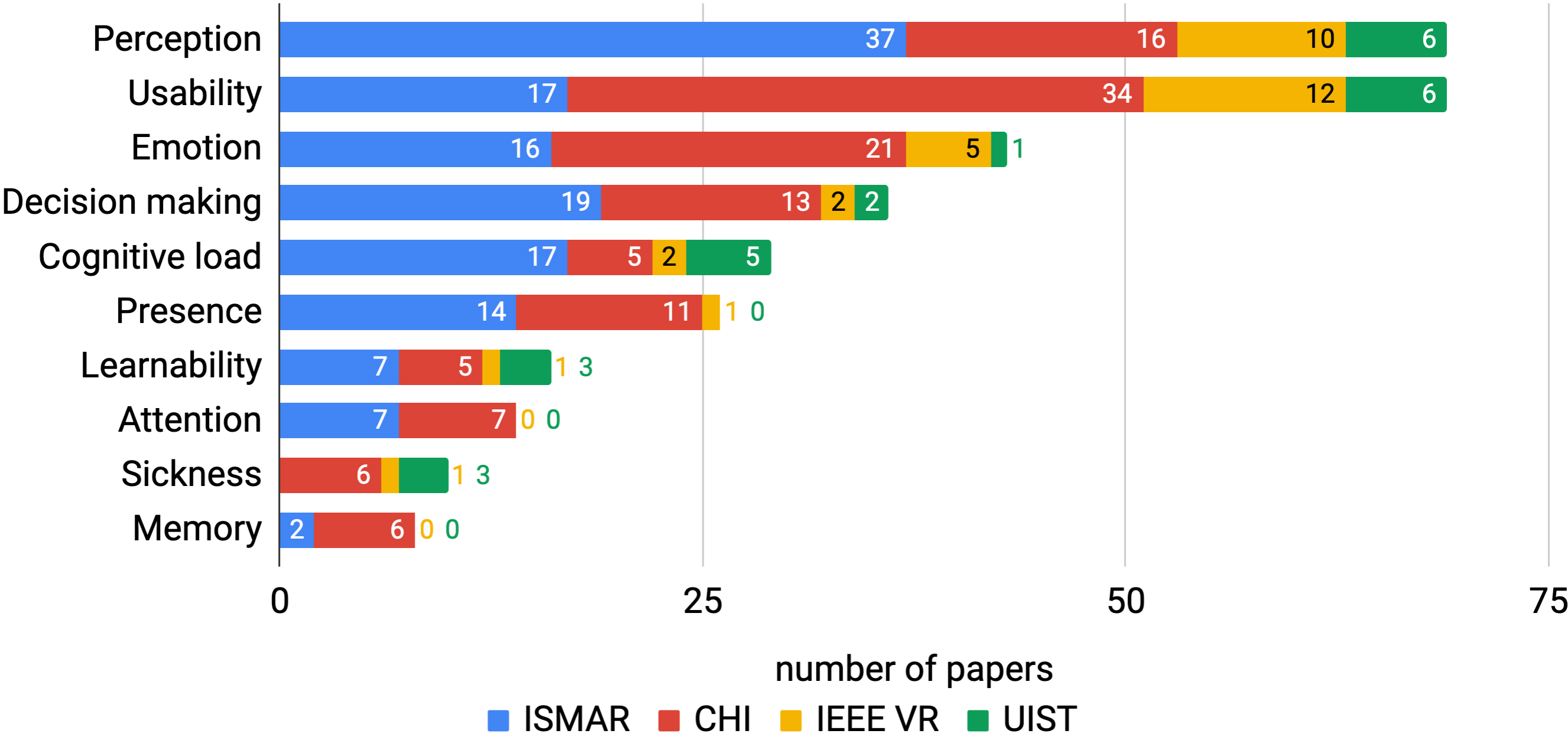}
  \caption{The number of papers that involve various cognitive aspects in MR/AR evaluation.}
  \label{fig:cognitive_aspects}
\end{figure}

We present the list of the 43 methods in~\autoref{tab:methods}, and a summary of the 10 inferred cognitive aspects in~\autoref{fig:cognitive_aspects}.
Notice that cognitive aspects (and data collection methods) are not mutually exclusive. That is, a user evaluation can involve multiple of them.
In particular, 
28\% of user evaluations (69/248) involve various aspects of perception (\eg visual, haptic, auditory), and a similar number of evaluations examined the usability of MR/AR approaches.
We found that 17\% of evaluations (43/248) involve the analysis of emotions, \eg intuitiveness, usefulness, or joyfulness. 
We rarely found decision making explicitly mentioned as an aspect of evaluations in MR/AR papers. However, we identified 15\% of user evaluations (36/248) that implicitly focus on it. Usually, these evaluations target MR/AR approaches that support users to make better decisions, for instance, in a short time and with high accuracy. 
Cognitive load is a frequent aspect involved in 12\% of user evaluations (29/248). 
Presence is included in 10\% of evaluations (26/248), which sometimes are combined with the analysis of awareness, embodiment, discernability, immersion, influence, and privacy.
Less frequently, we found evaluations that examined 
\begin{inparaenum}[(a)]
    \item learnability (16/248) by means of pre- and post-tests of performance, 
    \item attention (14/248), typically by means of eye-tracking technology,
    \item motion sickness (10/248), usually to assess fatigue amongst participants, and
    \item memory (8/248), regarding learning and spatial memory.
\end{inparaenum}
We also found a few other papers that reflect on cognitive aspects (but not in the context of an evaluation). We found two surveys: one~\cite{6162874} that reports on emotions of end-users who adopted an MR/AR tool, and another~\cite{8115401} that describes collected data for the analysis of user perception in the context of MR/AR. 
We found two model papers. One of them~\cite{6402561} discusses the impact of MR/AR in cognitive aspects (\ie perception, attention, memory). The other model paper~\cite{5643530} presents a taxonomy to characterize human perception.
We found three highly comprehensive MR/AR evaluations~\cite{3300458,3300774,3300674} that involved five cognitive aspects and three evaluations~\cite{8456570,3300431,3300403} that involve four cognitive aspects each. We also found that 14 papers described evaluations that involve 3 cognitive aspects, and the remaining 228 user studies involved up to 2 cognitive aspects.

We now describe examples of data collection methods of each inferred cognitive aspect. 

\begin{table*}[h!]
\caption{Data Collection Methods.}
\centering
\label{tab:methods}
\setlength\tabcolsep{3pt}
\renewcommand\arraystretch{0.25}
\begin{tabular}{lllllll} \\ \toprule
\textbf{Aspect} & \textbf{Method} & \textbf{Description} & \textbf{Ref.} &\textbf{Approach} & \textbf{Type} & \textbf{Sense}\\ \midrule
Perception &\textbf{2-AFC}& Two-Alternative Forced-Choice method&\cite{Ferw08a}&Obj. & Quant.&A/H/V \\ 
&\textbf{2-IFC}& Two-Interval Forced-Choice method&\cite{Yesh08a}&Obj. & Quant.&A/H/V \\
&\textbf{ACR11-HR}& Absolute Category Rating&\cite{Tomi10a,Itu99a}&Subj.  & Both.&V \\ 
&\textbf{SAQI}& Spatial Audio Quality Inventory&\cite{Lind14a}&Subj.  & Quali.&A \\[3pt]
Usability&\textbf{AD3}& Ad-hoc Usability Questionnaire (Awareness)&\cite{8613752}&Subj.  & Quant.&\\
&\textbf{MREQ}& Mixed Reality Experience Questionnaire&\cite{Rege17a}&Subj.  & Quant.&  \\
&\textbf{PEQ}& Post Experience Questionnaire&\cite{Llob13a}&Subj.  & Quant.&  \\
&\textbf{SUS}& The Slater-Usoh-Steed Questionnaire&\cite{Slat94a}&Subj.  & Quant.&  \\[3pt]
Emotion &\textbf{GEQ}& Game Experience Questionnaire&\cite{Ijss08a}&Subj.  & Quant.& \\ 
&\textbf{IMI}& Intrinsic Motivation Inventory&\cite{Mcau89a}&Subj.  & Quant.& \\
&\textbf{PANAS}& Positive and Negative Affect Schedule&\cite{Wats88a}&Subj.  & Quant.& \\
&\textbf{SAM}& Self-Assessment Manikin&\cite{Brad94a}&Subj.  & Quant.& \\
&\textbf{USQ}& IBM's Usability Satisfaction Questionnaire&\cite{Lewi95a}&Subj.  & Quant.&\\[3pt]
Presence &\textbf{AD1}& Ad-hoc Post Experimental Questionnaire&\cite{6948430}&Subj.  & Quant.&  \\ 
&\textbf{AD2}& Ad-hoc Co-Presence Questionnaire&\cite{6948412,7523400}&Subj.  & Quant.& \\  
&\textbf{BRQ}& Body Representation Questionnaire (Embodiment)&\cite{Bana13a}&Subj.  & Quant.&\\
&\textbf{IOS}& Inclusion of Other in the Self Scale&\cite{Aaro92a}&Subj.  & Quant.&  \\
&\textbf{IPQ}& The Igroup Presence questionnaire&\cite{Schu01a}&Subj.  & Quant.&  \\
&\textbf{MEC}& Spatial Presence Questionnaire&\cite{Vord04a}&Subj.  & Quant.&  \\
&\textbf{SPQ}& Social Presence Questionnaire&\cite{Harm19a}&Subj.  & Quant.&  \\
&\textbf{MTQ}& McKnight Trust Questionnaire (Trust)&\cite{Mckn11a}&Subj.  & Quant.&  \\
&\textbf{TPI}& The Temple Presence Inventory&\cite{Lomb09a}&Subj.  & Quant.& \\[3pt]
Cognitive load &\textbf{NASA-TLX}& NASA-TLX (Task Load Index)&\cite{Hart88a}&Subj. & Quant.&\\ 
&\textbf{SMEQ}& Subjective Mental Effort Questionnaire&\cite{Zijl93a}&Subj. & Quant. &\\
&\textbf{Paas}& Paas Mental-Effort Rating Scale&\cite{Paas92b}&Subj.  & Quant.&\\
&\textbf{ECG}& Electrocardiogram&\cite{5643560}&Obj.  & Quant.&\\
&\textbf{GSR}& Galvanic Skin Response&\cite{5643560}&Obj.  & Quant.&\\
&\textbf{ST}& Skin Temperature&\cite{5643560}&Obj.  & Quant.&\\ [3pt]
Attention &\textbf{2-AFC} & Two-Alternative Forced-Choice method&\cite{Ferw08a,Adam71a}&Subj.  & Quant.&V/A \\ 
&\textbf{AD4} & Ad-hoc Self-Report Questionnaire&\cite{8613754}&Subj.  & Quant.&A\\ 
&\textbf{VisEng.} & User Engagement Self-Report Questionnaire&\cite{8613760}&Subj.  & Quant.&A\\ 
&\textbf{ET} & Eye-tracking&\cite{8115412,7164337,8457524}&Obj.  & Quant.&V\\ 
&\textbf{HT} & Head-tracking&\cite{8115412,8457524}&Obj.  & Quant.&V\\[3pt]
Learnability &\textbf{PRE}& Pre-tests&\cite{8115415}&Subj.  & Both&  \\ 
&\textbf{POS}& Post-tests&\cite{7781773}&Subj.  & Both&  \\ 
&\textbf{PAS}& Pattern-of-Search&\cite{5336485}&Obj.  & Both&  \\ 
&\textbf{LOP}& Level of Pressure&\cite{5336485}&Obj.  & Quant.& \\ 
&\textbf{VAK}& Learning Styles Self-Report Questionnaire&\cite{Chis05a}&Subj.  & Both& \\[3pt]
Motion sickness&\textbf{SSQ}& Simulator Sickness Questionnaire (Fatigue)&\cite{Kenn03a}&Subj. & Quant.&\\[3pt]
Memory &\textbf{ASC}& Awareness State score&\cite{Coxo14a}&Subj.  & Quant.& \\ 
&\textbf{ANAM}& Automated Neuropsychological Assessment Metrics&\cite{Kane07a}&Subj.  & Quant.& \\ 
&\textbf{CUED}& Cue recall&\cite{8462799}&Subj.  & Both& \\ 
&\textbf{FREE}& Free recall&\cite{8462799}&Subj.  & Both& \\ 
\bottomrule
\end{tabular}
\end{table*}

\begin{table*}[!ht]
\caption{A summary of research topics in MR/AR by paper type, evaluation scenario, cognitive aspect, and configuration.}
\label{tab:research_topics}
\centering
\setlength\tabcolsep{0.2pt}
\renewcommand\arraystretch{0.1}
\begin{tabular}{lp{3pt}KKKKKp{3pt}rp{8pt}NNNNNNNp{8pt}MMMMMMMMMMp{8pt}YYYY} \\
 & & \multicolumn{1}{c}{\parbox[t]{2mm}{\rotatebox[origin=l]{90}{Technique}}} & \multicolumn{1}{c}{\parbox[t]{2mm}{\rotatebox[origin=l]{90}{Evaluation}}} & \multicolumn{1}{c}{\parbox[t]{2mm}{\rotatebox[origin=l]{90}{Design Study}}}& \multicolumn{1}{c}{\parbox[t]{2mm}{\rotatebox[origin=l]{90}{System}}}& \multicolumn{1}{c}{\parbox[t]{2mm}{\rotatebox[origin=l]{90}{Model}}} & & \multicolumn{1}{c}{\parbox[t]{2mm}{\rotatebox[origin=l]{90}{Total}}} & & \multicolumn{1}{c}{\parbox[t]{2mm}{\rotatebox[origin=l]{90}{AP}}}&\multicolumn{1}{c}{\hspace{1pt}\parbox[t]{2mm}{\rotatebox[origin=l]{90}{QRI}}}&\multicolumn{1}{c}{\hspace{3pt}\parbox[t]{2mm}{\rotatebox[origin=l]{90}{UE}}}& \multicolumn{1}{c}{\hspace{2pt}\parbox[t]{2mm}{\rotatebox[origin=l]{90}{UP}}}&\multicolumn{1}{c}{\hspace{3pt}\parbox[t]{2mm}{\rotatebox[origin=l]{90}{COL}}}&\multicolumn{1}{c}{\hspace{3pt}\parbox[t]{2mm}{\rotatebox[origin=l]{90}{COM}}}&\multicolumn{1}{c}{\hspace{3pt}\parbox[t]{2mm}{\rotatebox[origin=l]{90}{UWP}}} & & \multicolumn{1}{c}{\parbox[t]{2mm}{\rotatebox[origin=l]{90}{Perception}}} & \multicolumn{1}{c}{\parbox[t]{2mm}{\rotatebox[origin=l]{90}{Usability}}} &
  \multicolumn{1}{c}{\parbox[t]{2mm}{\rotatebox[origin=l]{90}{Emotion}}} & 
  \multicolumn{1}{c}{\parbox[t]{2mm}{\rotatebox[origin=l]{90}{Decision making}}}  & 
  \multicolumn{1}{c}{\parbox[t]{2mm}{\rotatebox[origin=l]{90}{Cognitive load}}} & 
  \multicolumn{1}{c}{\parbox[t]{2mm}{\rotatebox[origin=l]{90}{Presence}}} & 
  \multicolumn{1}{c}{\parbox[t]{2mm}{\rotatebox[origin=l]{90}{Learning}}} & 
  \multicolumn{1}{c}{\parbox[t]{2mm}{\rotatebox[origin=l]{90}{Attention}}}  &
  \multicolumn{1}{c}{\parbox[t]{2mm}{\rotatebox[origin=l]{90}{Motion sickness}}} & 
  \multicolumn{1}{c}{\parbox[t]{2mm}{\rotatebox[origin=l]{90}{Memory}}} & & 
  \multicolumn{1}{c}{\parbox[t]{2mm}{\rotatebox[origin=l]{90}{Static\textbackslash{}Lab}}} &
  \multicolumn{1}{c}{\parbox[t]{2mm}{\rotatebox[origin=l]{90}{Mobile\textbackslash{}Lab}}} & 
  \multicolumn{1}{c}{\parbox[t]{2mm}{\rotatebox[origin=l]{90}{Mobile\textbackslash{}In-the-Wild}}} & 
  \multicolumn{1}{c}{\parbox[t]{2mm}{\rotatebox[origin=l]{90}{Static\textbackslash{}In-the-Wild}}} \\ 
\multicolumn{2}{c}{\textbf{Topic}}&\multicolumn{7}{c}{\textbf{Paper Type}}&\multicolumn{8}{c}{\textbf{Evaluation Scenario}}& \multicolumn{11}{c}{\textbf{Cognitive Aspect}}& \multicolumn{5}{c}{\textbf{Configuration}}\\
  \toprule
Design/human factors & & 11 & 69 & 8 & 0 & 1 & & 89 &  & 7 & 2 & 57 & 57 & 6 & 4 & 1 & & 23 & 14 & 13 &7 & 15 & 11 & 7 & 4 & 3 & 6& & 43 & 32 & 9 & 1 \\
Tracking & & 65 & 3 & 0 & 1 & 0 & & 69 &  & 62 & 49 & 4 & 7 & 1 & 0 & 0 & & 3 & 2 & 0 &3 & 0 & 0 & 0 & 0 & 0 & 0& & 2 & 4 & 0 & 0 \\
Reconstruction & & 36 & 1 & 1 & 4 & 0 & & 42 &  & 30 & 33 & 5 & 5 & 0 & 0 & 0 & & 2 & 1 & 1 &0 & 2 & 0 & 2 & 0 & 1 & 0& & 4 & 4 & 0 & 0 \\
Multimodal interfaces & & 17 & 12 & 7 & 1 & 0 & & 37 &  & 6 & 7 & 21 & 23 & 0 & 1 & 1 & & 14 & 7 & 5 &7 & 3 & 3 & 0 & 2 & 1 & 0& & 24 & 6 & 2 & 0 \\
Applications & & 15 & 15 & 2 & 2 & 0 & & 34 &  & 9 & 9 & 20 & 15 & 0 & 1 & 4 & & 1 & 7 & 5 &8 & 3 & 0 & 4 & 1 & 0 & 1& & 8 & 4 & 8 & 1 \\
Displays & & 32 & 1 & 0 & 1 & 0 & & 34 &  & 23 & 23 & 8 & 8 & 1 & 1 & 0 & & 6 & 1 & 2 &1 & 0 & 1 & 0 & 1 & 0 & 0& & 8 & 2 & 0 & 0 \\
HCI technologies & & 20 & 12 & 1 & 1 & 0 & & 34 &  & 9 & 7 & 21 & 22 & 2 & 1 & 0 & & 3 & 6 & 3 &5 & 1 & 3 & 0 & 1 & 1 & 0& & 16 & 7 & 3 & 0 \\
Rendering & & 26 & 1 & 0 & 0 & 0 & & 27 &  & 19 & 23 & 7 & 6 & 0 & 1 & 0 & & 4 & 2 & 0 &1 & 0 & 0 & 0 & 0 & 0 & 0& & 6 & 2 & 0 & 0 \\
Collaboration & & 5 & 9 & 6 & 2 & 0 & & 22 &  & 4 & 2 & 17 & 7 & 14 & 7 & 0 & & 1 & 5 & 4 &0 & 1 & 7 & 0 & 1 & 1 & 1& & 9 & 9 & 2 & 0 \\
Calibration/registration & & 17 & 0 & 0 & 0 & 0 & & 17 &  & 15 & 10 & 1 & 2 & 0 & 0 & 0 & & 0 & 1 & 0 &0 & 0 & 0 & 0 & 0 & 0 & 0& & 0 & 1 & 0 & 0 \\
Mediated reality & & 9 & 1 & 2 & 2 & 0 & & 14 &  & 5 & 4 & 7 & 8 & 1 & 0 & 0 & & 4 & 1 & 0 &1 & 0 & 1 & 1 & 1 & 2 & 0& & 7 & 2 & 0 & 0 \\
Spatial annotation & & 11 & 1 & 1 & 1 & 0 & & 14 &  & 6 & 2 & 8 & 10 & 0 & 1 & 0 & & 2 & 4 & 2 &1 & 4 & 0 & 2 & 1 & 1 & 0& & 5 & 3 & 3 & 0 \\
Diminished reality & & 8 & 1 & 0 & 0 & 0 & & 9 &  & 5 & 8 & 0 & 4 & 0 & 0 & 0 & & 3 & 0 & 0 &1 & 0 & 0 & 0 & 0 & 0 & 0& & 4 & 0 & 0 & 0 \\
Data visualization & & 3 & 2 & 2 & 0 & 0 & & 7 &  & 0 & 0 & 5 & 5 & 1 & 2 & 0 & & 2 & 1 & 0 &1 & 0 & 0 & 0 & 2 & 0 & 0& & 2 & 2 & 3 & 0 \\
Taxonomy & & 0 & 2 & 0 & 0 & 5 & & 7 &  & 0 & 0 & 2 & 0 & 0 & 0 & 4 & & 1 & 1 & 1 &0 & 0 & 0 & 0 & 0 & 0 & 0& & 0 & 0 & 0 & 0 \\
Software architecture & & 2 & 0 & 0 & 0 & 0 & & 2 &  & 2 & 0 & 0 & 0 & 0 & 0 & 0 & & 0 & 0 & 0 &0 & 0 & 0 & 0 & 0 & 0 & 0& & 0 & 0 & 0 & 0 \\
\cmidrule[0.1pt]{1-33}
 \multicolumn{1}{l}{\textbf{Total}} & 
&\multicolumn{1}{r}{\parbox{1.8em}{277}}
&\multicolumn{1}{r}{\parbox{1.8em}{130}}
&\multicolumn{1}{r}{\parbox{1.3em}{\hfill 30}}
&\multicolumn{1}{r}{\parbox{1.3em}{\hfill 15}}
&\multicolumn{1}{r}{\parbox{1.3em}{\hfill 6}} &
&\multicolumn{1}{r}{\parbox{1.8em}{458}}& 
&\multicolumn{1}{r}{\parbox{1.8em}{202}}
&\multicolumn{1}{r}{\parbox{1.8em}{179}}
&\multicolumn{1}{r}{\parbox{1.8em}{183}}
&\multicolumn{1}{r}{\parbox{1.8em}{179}}
&\multicolumn{1}{r}{\parbox{1.3em}{\hfill 26}}
&\multicolumn{1}{r}{\parbox{1.3em}{\hfill 19}} 
&\multicolumn{1}{r}{\parbox{1.3em}{\hfill 10}} &  
&\multicolumn{1}{r}{\parbox{1.3em}{\hfill 69}}
&\multicolumn{1}{r}{\parbox{1.3em}{\hfill 53}} 
&\multicolumn{1}{r}{\parbox{1.3em}{\hfill 36}}  
&\multicolumn{1}{r}{\parbox{1.3em}{\hfill 36}} 
&\multicolumn{1}{r}{\parbox{1.3em}{\hfill 29}}
&\multicolumn{1}{r}{\parbox{1.3em}{\hfill 26}}
&\multicolumn{1}{r}{\parbox{1.3em}{\hfill 16}}
&\multicolumn{1}{r}{\parbox{1.3em}{\hfill 14}}
&\multicolumn{1}{r}{\parbox{1.3em}{\hfill 10}} 
&\multicolumn{1}{r}{\parbox{1.3em}{\hfill 8}} &
&\multicolumn{1}{r}{\parbox{1.8em}{ 138}}
&\multicolumn{1}{r}{\parbox{1.3em}{\hfill 78}}
&\multicolumn{1}{r}{\parbox{1.3em}{\hfill 30}}
&\multicolumn{1}{r}{\parbox{1.3em}{\hfill 2}} \\
\bottomrule
\end{tabular}
\end{table*}

\vspace{0.2em}
\noindent\textbf{Perception. }
We found that 28\% of user studies (69/248) analyze perception to examine topics such as design/human factors (23/69) and multimodal interfaces (14/69), which employ various methods depending on the type of perception.

\emph{Visual. }
When conducting perception studies, researchers selected either objective data collection methods, such as tracking head, eye, and body movements, or subjective data collection methods, such as 
\begin{inparaenum}[(a)]
    \item Absolute Category Rating (ACR11-HR)~\cite{7523376} (also called Single Stimulus Method). In it, participants are asked to evaluate the quality of a sequence of images that are presented one at a time and rated independently on a category scale. 
    \item Two-Alternative Forced-Choice (2-AFC)~\cite{8466636} has been used to measure various types of perception and attention. In it, participants are required to perform a central task and a peripheral task (\eg based on the visual angle or spatial location) simultaneously. For instance, researcher asked participants to scan a display panel, while at the same time, participants had to respond to light stimuli perceived in the periphery of their visual field. Head and eye movements are sometimes restricted depending on the focus of the evaluation. 
\end{inparaenum}    
\emph{Haptic. }
Studies that focus on haptic perception (\eg softness or stiffness) have used the Two-Interval Forced-Choice (2-IFC) method~\cite{5336501}, which is similar to 2-AFC, 
but in which options are presented sequentially in two intervals. The studies analyzed two variables:
\begin{inparaenum}[(a)]
    \item Just Noticeable Difference (JND), which is the point at which participants do not perceive differences between two similar options, and
    \item Point of Subjective Equality (PSE), which is the point at which participants perceive options of different nature as equal.
\end{inparaenum}
\emph{Auditory. }
Studies that analyzed, in particular, the relationship between auditory perception and spatial perception employed the Spatial Audio Quality Inventory (SAQI) vocabulary~\cite{8613754}, which is intended for a qualitatively comparative auditory assessment of acoustic scenes. 

\vspace{0.2em}
\noindent\textbf{Usability.}
We found that 21\% of user studies (53/248) assessed the usability of MR/AR approaches. These studies often focused on design/human factors (14/53) and involved methods such as:
\begin{inparaenum}[(a)]
    \item The Slater-Usoh-Steed Questionnaire (SUS)~\cite{8462799,8456525,8115413} ,
    \item Mixed Reality Experience Questionnaire (MREQ)~\cite{8115408}, and 
    \item Post Experience Questionnaire (PEQ)~\cite{8115408} .
\end{inparaenum}

\vspace{0.2em}
\noindent\textbf{Emotions. }
We found that 15\% of user studies (36/248) examined the emotions of participants.
We did not identify studies that applied objective methods to analyze emotions. Instead, we observed that studies often collect data of the emotions perceived by participants to examine various emotions using methods such as: 
\begin{inparaenum}[(a)]
    \item Game Experience Questionnaire (GEQ)~\cite{8115413} to measure game experience based on user engagement \eg competence, sensory and imaginative immersion, flow, challenge, positive affect, negative affect, tension, and annoyance;
    \item Intrinsic Motivation Inventory (IMI)~\cite{6671766} to measure the overall user experience with regard to general experimental tasks.
\end{inparaenum}
Other methods used to assess emotions in general were: 
\begin{inparaenum}[(a)]
    \item Positive and Negative Affect Schedule (PANAS)~\cite{8613762,8943737} to assess positive and negative affect and 
    \item Self-Assessment Manikin (SAM)~\cite{8613762,8943737}, which is a pictorial assessment technique to measure pleasure, arousal, and dominance associated with a person's affective impressions.
\end{inparaenum}

\vspace{0.2em}
\noindent\textbf{Decision making. }
We found that 15\% of user studies (36/248) analyzed decision making.
Such studies complemented an analysis of the time and the correctness of participants to complete tasks by collecting data of head movements (\eg rotation, exertion, or velocity) to examine effort, efficiency, and effectiveness~\cite{5336486,6671762}. 

\vspace{0.2em}
\noindent\textbf{Cognitive load. }
We found that 12\% of user studies (29/248) analyzed the cognitive load of participants exposed to MR/AR approaches.
Cognitive load involves two concepts: 
\begin{inparaenum}[(a)]
    \item mental load imposed by instructional parameters, \eg task structure, the sequence of information given during an evaluation, and 
    \item mental effort that refers to the capacity allocated by participants of a study to the instructional demands. 
\end{inparaenum}
Therefore, when evaluating cognitive load in laboratory settings, the mental load can be fixed and kept the same across the evaluated conditions. Thus, measures of mental effort can be considered an index of cognitive load.
Studies that examined cognitive load used objective data collection methods, for instance, to examine the anxiety of participants. There exist several physiological measures that have been examined to analyze mental effort, such as pupil dilation~\cite{Kahn73a}, heart rate variability~\cite{Muld80a}, event-related brain potentials~\cite{Donc81a}, muscle tension~\cite{Boxt93a}, adrenaline level~\cite{Fran73a}, skin temperature and galvanic skin response (GSR)~\cite{5643560}.
Studies also used subjective methods such as: 
\begin{inparaenum}[(a)]
    \item Paas 9-step mental-effort Likert scale (Paas)~\cite{8007333,8493594,8456525}, in which score indexes of metal effort go from ``very, very low effort'' (1) to ``very, very high effort'' (9);
    \item Subjective Mental Effort Question (SMEQ)~\cite{8613761}, in which participants indicate their mental effort using a scale that goes from ``not at all hard to do'' (0) to ``tremendously hard to do'' (150); and
    \item NASA-TLX~\cite{6948411,8613752,5336484} that is the assessment of total workload divided into six subscales: mental demand, physical demand, temporal demand, performance effort, and frustration.
\end{inparaenum}

\vspace{0.2em}
\noindent\textbf{Presence. }
We found that 10\% of user studies (26/248) analyzed presence.
Studies~\cite{5643560} that examine the feeling of presence use several objective methods to measure presence-based physiological responses: 
\begin{inparaenum}[(a)]
    \item Electrocardiogram (ECG),
    \item Galvanic Skin Response (GSR),
    \item skin temperature,
    \item brain activity (\eg EEG), 
    \item heart rate, and
    \item respiration rate.
\end{inparaenum}

Studies can also use several subjective data collection methods. The methods are based on questionnaires that participants in a study are asked to fill to cover various aspects of presence \eg co-presence, spatial presence, social presence, social richness, closeness, or connectedness.
Other methods that can be used to complement the analysis of presence are:
\begin{inparaenum}[(a)]
    \item Ad-Hoc Usability Questionnaire (AD3)~\cite{8613752} to analyze the \emph{awareness} of participants in an immersive environment; 
    \item Body Representation Questionnaire (BRQ)~\cite{8115408} for the analysis of embodiment;
    \item McKnight Trust Questionnaire (MTQ)~\cite{8613756} to assess whether participants trust in technology, \eg reliability, helpfulness, functionality, and situational normality.
\end{inparaenum}

 \emph{Simulator Sickness Questionnaire (SSQ)}~\cite{6671765,8466636,8613752}. SSQ is a well-known test to check symptoms of nausea, fatigue, and disorientation, which could affect the integrity of participants, and in consequence, the results of the evaluation. SSQ can be a suitable complement to the analysis of multiple cognitive aspects. However, we often observed its application in the assessment of presence.
 
\vspace{0.2em}
\noindent\textbf{Learnability. }
We found that 7\% of user studies (16/248) analyzed learnability promoted by MR/AR approaches.
Studies focused on learnability used general methods such as pre-tests to assess the prior knowledge of participants of a subject. The results offered researchers a baseline for comparing to the results of a post-test. A significant increase in the measured knowledge suggested that the approach under analysis promoted learnability. Other specific objective methods employed in the analysis of learnability of particular topics are
\begin{inparaenum}[(a)]
    \item Level of Pressure (LOP)~\cite{5336485} to assess learner’s use of correct pressure or 
    \item Pattern-of-Search (PAS)~\cite{5336485} that is applicable, for instance, to medical training of breast exams.
\end{inparaenum}
The study of learnability was complemented employing a classification the learning styles self-assessment questionnaire (VAK)~\cite{7781773}. Learning styles are characterized based on 
\begin{inparaenum}[(a)]
    \item the use of seen (visual),  
    \item the transfer of information through listening (auditory), and
    \item physical experience (kinesthetic).
\end{inparaenum}
However, we notice that the method has some detractors~\cite{sharp2008vak}.

\vspace{0.2em}
\noindent\textbf{Attention. }
We found that 6\% of user studies (14/248) analyzed the attention of participants of an MR/AR evaluation.
Studies that involved perception used various methods depending on the type of perception.

\emph{Visual. }
Studies selected objective methods to collect data from eye- and head-tracking. Using eye-tracking, experimenters analyzed when participants were distracted. Eye-tracking complemented with an analysis of head movements (\eg orientation angle of participants heads) indicated when participants were distracted as well. Visual attention was also analyzed based on subjective methods that consider, for instance, the assessment of engagement, through
\begin{inparaenum}[(a)]
    \item Ad-hoc Self-Report Questionnaire of Engagement (AD4)~\cite{8613754} or 
    \item User Engagement Self-Report Questionnaire (VisEngage)~\cite{8613760}.
\end{inparaenum}
\emph{Auditory. } 
Studies examined, in particular, the connection between listener sensitivity in audio localization and experienced attention for immersion in MR. Subjective auditory attention was measured using an ad-hoc self-report questionnaire (AD4). 
\emph{Eyesight tests}. Snellen eye charts~\cite{7164337} were used to ensure the visual aptitude of participants of studies that involved visual attention.

\vspace{0.2em}
\noindent\textbf{Memory. }
We found that 3\% of user studies (8/248) analyzed the recollection of participants of an MR/AR evaluation.
Studies that examined memory ensured first that participants in an evaluation had a normal memory ability. To this end, they used the Automated Neuropsychological Assessment Metrics (ANAM) survey~\cite{8462799}. Other general methods for collecting data on the recollection of participants in evaluations of MR/AR were:
\begin{inparaenum}[(a)]
    \item Free Recall, in which participants were asked to tell a narrative from their recollection;
    \item Cued Recall, in which participants were asked questions to drive their recollection in order to recognize relevant points that were missing in their narrative.
\end{inparaenum}
Since the confidence of participants in their recollection varied sometimes, researchers complemented the analysis with measures of participants' level of confidence using the Awareness State score~\cite{8462799,Coxo14a}.

\subsection{Study Configuration}
\begin{figure}[t]
  \centering
  \includegraphics[width=\linewidth]{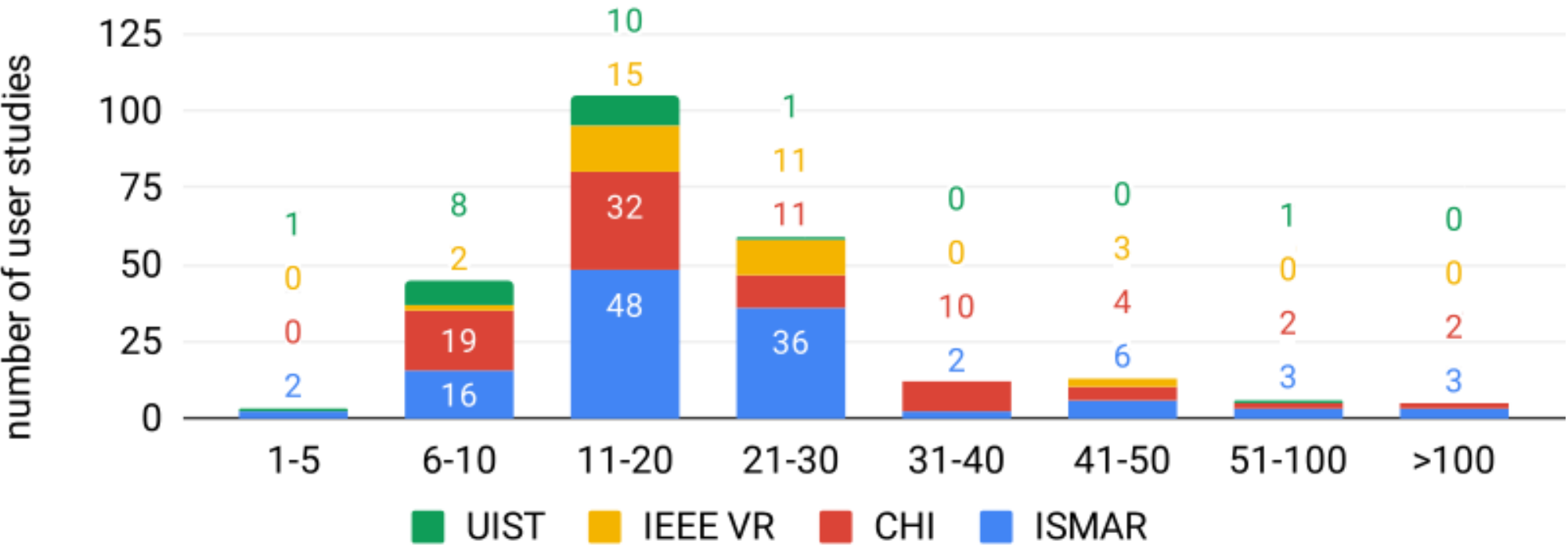}
  \caption{Histogram of the number of participants in user studies in MR/AR.}
  \label{fig:participants}
\end{figure}

We found that 54\% of all papers reported on evaluations with users (248/458). One important choice when designing a user evaluation is to define the number of participants that are going to be involved.
We found that user evaluations involved 5,761 participants in total, of which only 1,619 were identified as female and 3,087 as male.
\autoref{fig:participants} shows a histogram of the distribution of the samples sizes that we found among the reported user studies. 
We found that 93\% of user evaluations (231/248) include between 3 and 30 participants (median of 15) of which a median of 4 were females. The remaining 7\% of user evaluations include a median of 44 participants overall and a median of 16 female participants. 
The sample sizes in MR/AR seem larger than in HCI in general. A previous study~\cite{caine2016local} found that the most common sample size is 12 in evaluations published in CHI. Similarly, in visualization, most user studies involve ten or fewer participants~\cite{Isen13a}. We found that 42\% of the users studies (105/248) 
include 11--20 participants, 
and user studies that involve a smaller number of participants (\ie 1--5) are not frequent in MR/AR. Such smaller numbers are more common in qualitative studies, which do not seem to be as frequent in MR/AR as in other domains. 
The distribution of the number of participants involved in MR/AR user studies is  similar across the four analyzed venues.   
\begin{figure}[t]
  \centering
  \includegraphics[width=\linewidth]{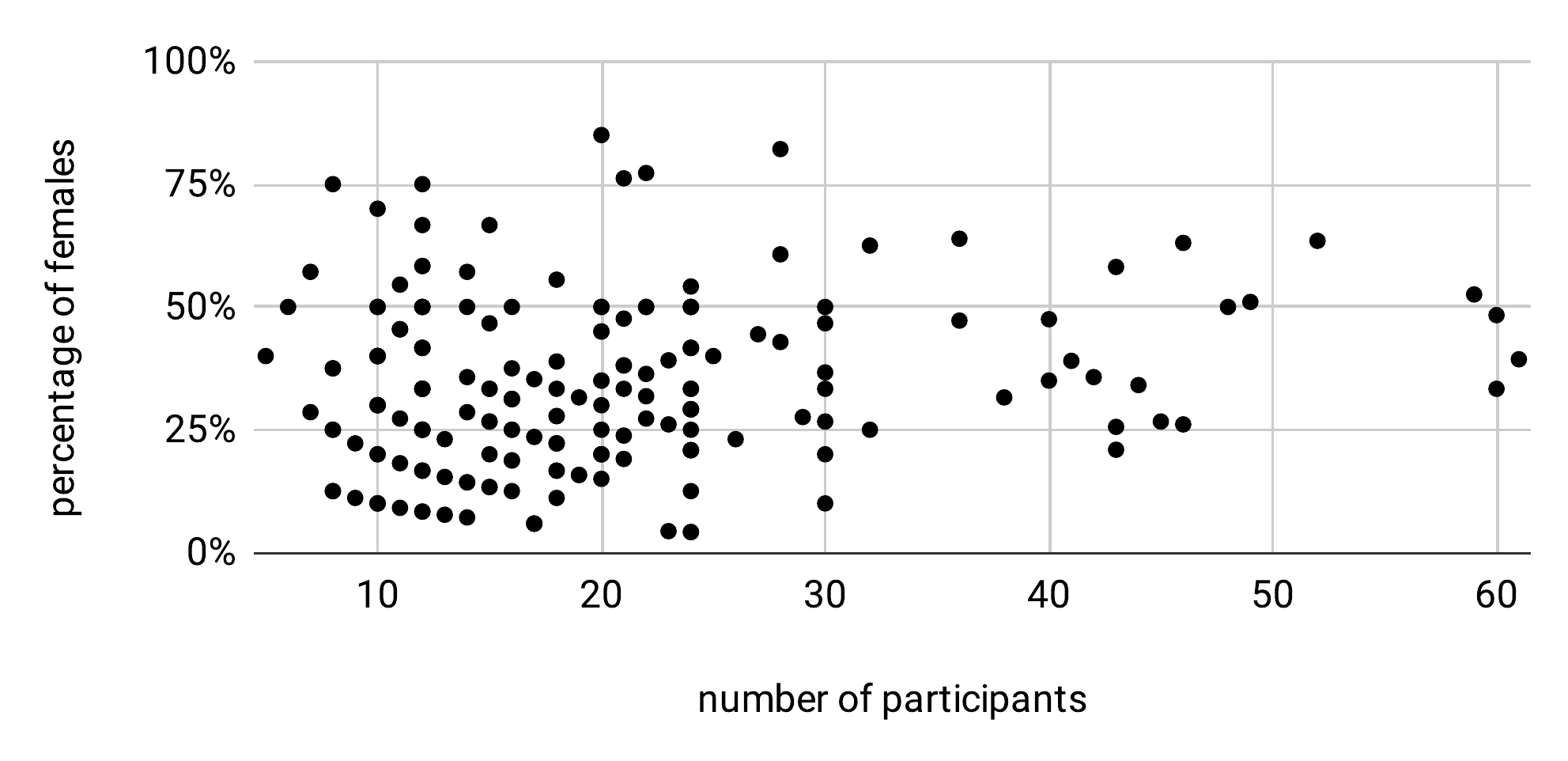}
  \caption{The sample sizes by gender of 204 of the 248 user evaluations in MR/AR (44 studies did not report on gender).}
  \label{fig:participants_gender}
\end{figure} 
\begin{figure}[t]
  \centering
  \includegraphics[width=\linewidth]{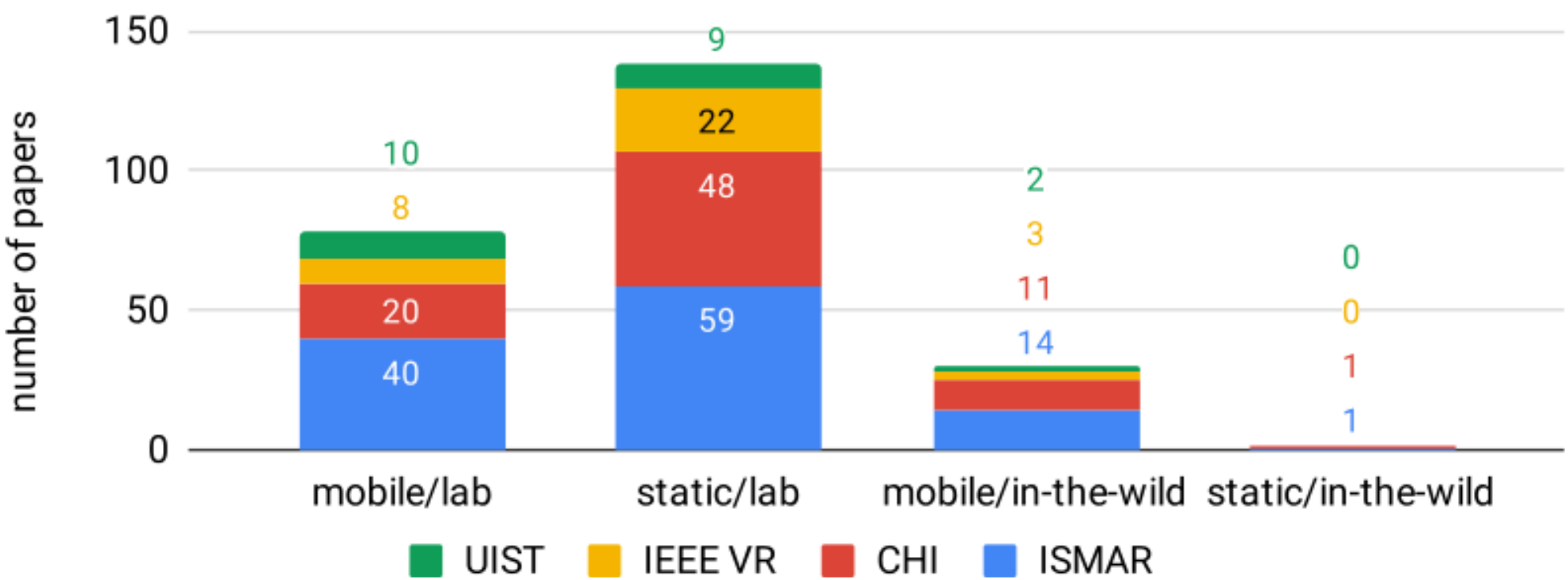}
  \caption{A classification of the 248 user studies in MR/AR per static/mobile and lab/in-the-wild configuration.}
  \label{fig:study_congiguration}
\end{figure}

We further examine the number of participants split by gender. We present in~\autoref{fig:participants_gender} a scatterplot with the the percentage of females versus the total number of participants. 
We observe that only a few studies involve a majority of female participants. 
We excluded from the chart 44 studies that do not specify the gender of participants.

We also analyzed  whether user studies were conducted in a laboratory setting or in-the-wild, as well as, whether participants in the study where static (\ie sitting or standing) or mobile. The results are shown in~\autoref{fig:study_congiguration}. We split the results by venue. We observe that the majority of user studies (\ie 82\% CHI--100\% IEEE VR) are conducted in a laboratory. Amongst such studies, in UIST, 50\% of user studies involve participants using MR/AR in a mobile way, whereas in IEEE VR, only 8\% do so. In contrast, user studies occasionally are conducted in-the-wild. User studies conducted in-the-wild usually involve participants in a mobile fashion, for example, to support needs in the automotive industry~\cite{8466859,8115412,6671764,8466859}. We only found two in-the-wild studies in which participants were static~\cite{7781773,Oppermann:2015:SPA:2702123.2702538}.
We did not identify in-the-wild studies published in IEEE VR. 












\section{Discussion and Guidance}
\label{sec:discussion}
We now discuss considerations of cross-cutting concerns that emerged from the analyzed dimensions that can guide researchers to design evaluations in MR/AR.

\vspace{0.2em}
\noindent\textbf{Bridging technology-centric and human-centric evaluations.} 
 We found two main types of MR/AR papers: 
\begin{inparaenum}[(a)]
    \item technology-centric papers (293/458) that focus on topics such as tracking, displays, reconstruction, rendering, or calibration, which are frequently evaluated through AP+QRI scenarios, and 
    \item human-centric papers (160/458) that deal with topics such as applications, design/human factors, which are frequently evaluated through UP+UE scenarios.
\end{inparaenum}
There seem to be a few small crossover topics, \ie spatial annotation, multimodal interfaces, or collaboration. 
However, technology-centric and human-centric approaches use methodologies that are mostly disjoint. Consequently, we ask how we can bridge technology-centric and human-centric evaluations when this combination helps address a research question. In fact, we have observed some papers that followed such pattern~\cite{5643564,7523389,7523376,7523388,8456571,2208641,2208297,3347933,8943618, 8943645, 8943651}.
We call researchers in the field to complement the results of benchmarks with user studies when techniques involve a user interface and the combination of evaluations contributes to the research question at hand.

\vspace{0.2em}
\noindent\textbf{Toward comprehensive cognitive methods.} 
We identified that 236 out of 297 cognitive aspects are involved in evaluations of design/human factors, multimodal interfaces, applications, HCI technologies, collaboration, and spatial annotation (see~\autoref{tab:methods}). Based on their frequency (see~\autoref{tab:research_topics}), major concerns for 
\begin{inparaenum}[(a)]
    \item design/human factors are perception, presence, and cognitive load,
    \item applications are decision making, emotions, and cognitive load/perception,
    \item multimodal interfaces are perception and emotions, and
    \item spatial annotations are cognitive load, presence, and emotions.
\end{inparaenum}
We observe that the wide range of inherent topics to MR/AR can pose a challenge for newcomers to the field. We think our paper can help newcomers with such directions.

We found 34 subjective and nine objective data collection methods (see~\autoref{tab:methods}). 
We observe that evaluations that included objective data collection methods usually reported difficulties encountered when analyzing and interpreting the collected data (\eg noisy data)~\cite{5643560} and represent a valuable source of knowledge. 
Although subjective questionnaires can be helpful to collect impressions of the perception of participants, we observe a need for objective data that avoids biases of subjective data collection methods. For instance, participants in a study might be biased to share positive rather than negative emotions. 
Objective data are not limited to physiological sources, \eg ECG, EEG, GSR, but can include behavioral measures as well, \eg eye-, head-, and body-tracking.   
We call researcher in the field to involve multiple cognitive aspects that are suitable to comprehensively examine a given research question.

\vspace{0.2em}
\noindent\textbf{Gender bias limitations due lack of female participants.} 
Although our results show that participants populations in MR/AR are larger than populations in studies in HCI~\cite{caine2016local} and visualization~\cite{Isen13a}, we found that often such populations exhibit unbalanced genders (see~\autoref{fig:participants_gender}).
Certainly, the unbalanced gender of participants in studies is related to the recruiting strategies, which often rely on the (predominantly male) student population in computer science programs. However, gender bias can be an important concern in multiple fields~\cite{Coir19a} and should not be taken lightly. Recruiting an adequate number of 
participants is a step to more credible quantitative evaluation, but it should not be done at the expense of introducing gender bias.
We call researchers in the field to involve an adequate participant population with a more inclusive and gender-balanced distribution.

\vspace{0.2em}
\noindent\textbf{Increasing the ecological validity of evaluations.} 
The papers that we classified in the design study type (30), the applications topic (34), the UWP scenario (10), or the in-the-wild configuration (32) 
report on MR/AR approaches that were successfully used to deal with concrete real-world cases. 
Studying a real-world use case in-situ is clearly challenging, but, with more mature enabling technology at our disposal, successful real-world deployment has become realistic. We call to researchers in the field to conduct case studies that can help investigate in-depth phenomena in a concrete real-world case. 
We found that 87\% of user evaluations (216/248) are conducted in a laboratory setting, of which 78 involve participants in a mobile fashion. We found that 13\% of studies (32/248) are conducted in-the-wild, which often involve participants in a mobile setting.
We call researchers in the field to increase ecological validity by conducting in-the-wild evaluations with mobile or static participants depending on the targeted user behavior.

\vspace{0.2em}
\noindent\textbf{Depth-first (re)search: qualitative vs. quantitative analysis. }
We observe that thorough evaluations that entail qualitative approaches (\eg case studies) can facilitate a deep understanding of a phenomenon~\cite{Carp08a}. 
However, we found that only 8 out of 43 different data collection methods used in MR/AR can support qualitative analysis (see~\autoref{tab:methods}).
Due to the nature of qualitative data, evaluations that adopt a qualitative approach usually involve a limited number of participants, and thus, the results of these evaluations are hard to generalize to a large population. In contrast, evaluations that use quantitative methods can involve a higher number of participants, and, through the use of statistical analysis, can generalize results. However, such studies need to be highly controlled, hampering the application of results under real-world circumstances (ecological validity). Picking the right method is a well-known trade-off process with different methods offering different benefits and drawbacks~\cite{mcgrath1995methodology}. 
We think that to formulate appropriate hypotheses that can be thoroughly tested and generalized using quantitative methods, first, researchers need to obtain a deep understanding of the examined phenomenon and identify which qualitative methods are suited best. 
An approach of depth-first and breadth-second (re)search is clearly underrepresented in the surveyed MR/AR work, and we speculate that it may offer interesting findings in several cases. 



\vspace{0.2em}
\noindent\textbf{Where does my paper fit best: ISMAR, CHI, IEEE VR, UIST?}
We now discuss the types of work that are mostly accepted at these venues. Certainly, ISMAR is the main venue for MR/AR. Whereas in the past most papers published in ISMAR centered on techniques, today papers mostly focus on human-centered evaluations. Usually, ISMAR papers describe thorough evaluations that involve the analysis of multiple cognitive aspects to help researchers address the analysis of complex phenomena. CHI exhibits an increasing trend of MR/AR papers with a balanced interest in techniques, evaluations, and, to a lesser degree, design studies. Although CHI papers mostly focus on design, human factors, and HCI technologies, there is also interest in research that focuses on multimodal interfaces and collaboration. Recently, IEEE VR and CHI doubled the number of MR/AR papers with a balanced number of techniques and evaluations. IEEE VR papers mostly focus on design and human factors. In UIST, there is a small and consistent number of MR/AR papers published every year, which mostly correspond to techniques. Often, technique papers published in UIST focus on reconstruction whereas evaluation papers focus on HCI technologies.

\vspace{0.2em}
\noindent\textbf{The future of MR/AR}. In the future, we expect that the number of MR/AR papers will keep increasing and remain balanced amongst ISMAR, CHI, and IEEE VR, and, to a lesser extent, UIST. As MR/AR technologies become more mature, questions that involve human aspects will gain focus in MR/AR research. Consequently, we expect that future MR/AR papers will elaborate on human-centered evaluations that involve not only the analysis of user performance and user experience, but also the analysis of other scenarios, like understanding the role of MR/AR in working places and in communication and collaboration. Hence, we envision that there will be an increasing need for developing methods that support researchers to deal with such scenarios, which might involve in-the-wild configurations.
Our results confirm that MR/AR is a very complex technology. We observe that, even in laboratory settings, it is difficult to conduct a user study in which higher-level cognition is tested without being confounded by imperfections of MR/AR technology. For example, it is very hard to perform long-term studies if mobile devices run out of batteries. Tasks that require much time will likely not be carried out completely. That could explain why authors tend to report on ``low-level'' user performance, which is often complemented with questionnaires like NASA TLX to document cognition aspects. The rise of commercial-grade devices like the Microsoft HoloLens lowers the barrier of having standardized conditions for evaluations, but we have yet to see this has an effect in publications.

\section{Conclusion}
\label{sec:conclusion}
We analyzed evaluations reported in 485 papers of the research literature of MR/AR. We confirmed that
\begin{inparaenum}[(a)]
    \item technology-centric evaluations (through benchmarks of tracking, displays, reconstruction, rendering, and calibration) and 
    \item human-centric evaluations (of applications, and design/human factors) are the core pillars of MR/AR evaluation.
\end{inparaenum}
We found a marginal number of team-centric evaluations that involve collaboration, communication, and understanding environments and work practices.
We call researchers in the field to conduct thorough evaluations by:
\begin{inparaenum}[(a)]
    \item conducting user studies that complement the results of benchmarks when techniques involve a user interface and the combination is coherent with a research question;
    \item involving multiple cognitive aspects that can help comprehensive examination of a research question; 
    \item choosing appropriate methods for assessing the impact of an approach in human cognition, for which they can consult our selected examples; 
    \item involving an adequate participant population with a more inclusive gender-balanced distribution; 
    \item increasing the ecological validity of evaluations through in-the-wild and mobile or static configurations depending on the intended user behavior.
\end{inparaenum}

\acknowledgments{Merino, Kraus, and Weiskopf acknowledge funding by the Deutsche Forschungsgemeinschaft (DFG, German Research Foundation) -- Project-ID 251654672 -- TRR 161. Schwarzl and Sedlmair were supported by the Deutsche Forschungsgemeinschaft (DFG, German Research Foundation) under Germany's Excellence Strategy -- EXC 2120/1 -- 390831618. 
}


\bibliographystyle{abbrv-doi}

\bibliography{hive,methods,chi,ismar}

\begin{thebibliography}{100}

\bibitem{acmdl}
{ACM DL}. https://dl.acm.org.
\newblock Accessed on 2020-05-01.

\bibitem{chi}
{CHI}. http://chi.acm.org/.
\newblock Accessed on 2020-05-01.

\bibitem{lexico}
{"Cognition". Lexico. https://www.lexico.com}.
\newblock Accessed on 2020-05-06.

\bibitem{core}
{CORE}. http://portal.core.edu.au/conf-ranks/.
\newblock Accessed on 2020-05-01.

\bibitem{ieeevis}
{IEEE VIS}. http://ieeevis.org/.
\newblock Accessed on 2020-05-01.

\bibitem{ieeevr}
{IEEE VR}. hhttp://ieeevr.org/.
\newblock Accessed on 2020-05-01.

\bibitem{ieeeexplor}
{IEEE Xplore}. https://ieeexplore.ieee.org.
\newblock Accessed on 2020-05-01.

\bibitem{ismar}
{ISMAR}. http://ismar.net/.
\newblock Accessed on 2020-05-01.

\bibitem{uist}
{UIST}. http://uist.acm.org.
\newblock Accessed on 2020-05-01.

\bibitem{Adam71a}
R.~W. Adams.
\newblock {\em Peripheral vision and visual attention}.
\newblock PhD thesis, Iowa State University, 1971.

\bibitem{Aaro92a}
A.~Aron, E.~Aron, and D.~Smollan.
\newblock Inclusion of other in the self scale and the structure of
  interpersonal closeness.
\newblock {\em Journal of Personality and Social Psychology}, 63:596--612, Oct.
  1992.

\bibitem{Bana13a}
D.~Banakou, R.~Groten, and M.~Slater.
\newblock Illusory ownership of a virtual child body causes overestimation of
  object sizes and implicit attitude changes.
\newblock {\em National Academy of Sciences}, 110(31):12846--12851, 2013.

\bibitem{8007333}
J.~{Baumeister}, S.~Y. {Ssin}, N.~A.~M. {ElSayed}, J.~{Dorrian}, D.~P. {Webb},
  J.~A. {Walsh}, T.~M. {Simon}, A.~{Irlitti}, R.~T. {Smith}, M.~{Kohler}, and
  B.~H. {Thomas}.
\newblock Cognitive cost of using augmented reality displays.
\newblock {\em Transactions on Visualization and Computer Graphics},
  23(11):2378--2388, 2017.

\bibitem{2208297}
D.~Baur, S.~Boring, and S.~Feiner.
\newblock Virtual projection: Exploring optical projection as a metaphor for
  multi-device interaction.
\newblock In {\em Proceedings of the ACM Conference on Human Factors in
  Computing Systems (CHI)}, pp. 1693–--1702, 2012.

\bibitem{8115415}
F.~{Bork}, R.~{Barmaki}, U.~{Eck}, K.~{Yu}, C.~{Sandor}, and N.~{Navab}.
\newblock Empirical study of non-reversing magic mirrors for augmented reality
  anatomy learning.
\newblock In {\em IEEE International Symposium on Mixed and Augmented Reality
  (ISMAR)}, pp. 169--176, 2017.

\bibitem{8456525}
F.~{Bork}, C.~{Schnelzer}, U.~{Eck}, and N.~{Navab}.
\newblock Towards efficient visual guidance in limited field-of-view
  head-mounted displays.
\newblock {\em Transactions on Visualization and Computer Graphics},
  24(11):2983--2992, 2018.

\bibitem{Brad94a}
M.~M. Bradley and P.~J. Lang.
\newblock Measuring emotion: The self-assessment manikin and the semantic
  differential.
\newblock {\em Journal of Behavior Therapy and Experimental Psychiatry},
  25(1):49--59, 1994.

\bibitem{6671765}
G.~{Bruder}, P.~{Wieland}, B.~{Bolte}, M.~{Lappe}, and F.~{Steinicke}.
\newblock Going with the flow: Modifying self-motion perception with
  computer-mediated optic flow.
\newblock In {\em IEEE International Symposium on Mixed and Augmented Reality
  (ISMAR)}, pp. 67--74, 2013.

\bibitem{caine2016local}
K.~Caine.
\newblock Local standards for sample size at {CHI}.
\newblock In {\em Proceedings of the ACM Conference on Human Factors in
  Computing Systems (CHI)}, pp. 981--992, 2016.

\bibitem{Carp08a}
S.~Carpendale.
\newblock Evaluating information visualizations.
\newblock In A.~Kerren, J.~T. Stasko, J.-D. Fekete, and C.~North, eds., {\em
  Information Visualization: Human-Centered Issues and Perspectives}, pp.
  19--45. Springer Berlin Heidelberg, Berlin, Heidelberg, 2008.

\bibitem{8115411}
L.~{Chen}, T.~W. {Day}, W.~{Tang}, and N.~W. {John}.
\newblock Recent developments and future challenges in medical mixed reality.
\newblock In {\em IEEE International Symposium on Mixed and Augmented Reality
  (ISMAR)}, pp. 123--135, 2017.

\bibitem{Itu99a}
J.-H. Choe, T.-U. Jeong, H.~Choi, E.-J. Lee, S.-W. Lee, and C.-H. Lee.
\newblock Subjective video quality assessment methods for multimedia
  applications.
\newblock {\em Journal of Broadcast Engineering}, 18(3):416--424, 1999.

\bibitem{8115413}
M.~A. {Cidota}, P.~J.~M. {Bank}, P.~W. {Ouwehand}, and S.~G. {Lukosch}.
\newblock Assessing upper extremity motor dysfunction using an augmented
  reality game.
\newblock In {\em IEEE International Symposium on Mixed and Augmented Reality
  (ISMAR)}, pp. 144--154, 2017.

\bibitem{Coir19a}
P.~Coiro and D.~D. Pollak.
\newblock Sex and gender bias in the experimental neurosciences: the case of
  the maternal immune activation model.
\newblock {\em Translational Psychiatry}, 9(1):90, 2019.

\bibitem{Coxo14a}
M.~Coxon and K.~Mania.
\newblock Measuring memories for objects and their locations in immersive
  virtual environments: The subjective component of memorial experience.
\newblock In {\em Handbook of Human Centric Visualization}, pp. 453--471.
  Springer, New York, NY, 2014.

\bibitem{Dey18a}
A.~Dey, M.~Billinghurst, R.~W. Lindeman, and J.~Swan.
\newblock A systematic review of 10 years of augmented reality usability
  studies: 2005 to 2014.
\newblock {\em Frontiers in Robotics and AI}, 5:37, 2018.

\bibitem{8943737}
A.~{Dey}, H.~{Chen}, A.~{Hayati}, M.~{Billinghurst}, and R.~W. {Lindeman}.
\newblock Sharing manipulated heart rate feedback in collaborative virtual
  environments.
\newblock In {\em IEEE International Symposium on Mixed and Augmented Reality
  (ISMAR)}, pp. 248--257, 2019.

\bibitem{8613762}
A.~{Dey}, H.~{Chen}, C.~{Zhuang}, M.~{Billinghurst}, and R.~W. {Lindeman}.
\newblock Effects of sharing real-time multi-sensory heart rate feedback in
  different immersive collaborative virtual environments.
\newblock In {\em IEEE International Symposium on Mixed and Augmented Reality
  (ISMAR)}, pp. 165--173, 2018.

\bibitem{Donc81a}
E.~Donchin.
\newblock Surprise!{\ldots} surprise?
\newblock {\em Psychophysiology}, 18(5):493--513, 1981.

\bibitem{Duen08a}
A.~Duenser, R.~Grasset, and M.~Billinghurst.
\newblock A survey of evaluation techniques used in augmented reality studies.
\newblock Technical report, Human Interface Technology Laboratory New Zealand,
  2008.

\bibitem{Elmq15a}
N.~Elmqvist and J.~S. Yi.
\newblock Patterns for visualization evaluation.
\newblock {\em Information Visualization}, 14(3):250--269, 2015.

\bibitem{Ferw08a}
J.~A. Ferwerda.
\newblock Psychophysics 101: how to run perception experiments in computer
  graphics.
\newblock In {\em ACM SIGGRAPH 2008 classes}, pp. 1--60. 2008.

\bibitem{6162889}
P.~{Fite-Georgel}.
\newblock Is there a reality in industrial augmented reality?
\newblock In {\em IEEE International Symposium on Mixed and Augmented Reality
  (ISMAR)}, pp. 201--210, 2011.

\bibitem{Chis05a}
N.~Fleming and D.~Baume.
\newblock Learning styles again: Varking up the right tree!
\newblock {\em Educational developments}, 7(4):4, 2006.

\bibitem{8770302}
A.~{Fonnet} and Y.~{Pri{\'{e}}}.
\newblock Survey of immersive analytics.
\newblock {\em Transactions on Visualization and Computer Graphics}, pp. 1--1,
  2019.

\bibitem{Fran73a}
M.~Frankenhaeuser.
\newblock Experimental approaches to the study of catecholamines and emotion.
\newblock In {\em Proceedings of the Symposium on Parameters of Emotion}, p.
  684–685, 1975.

\bibitem{6671764}
W.-T. Fu, J.~Gasper, and S.-W. Kim.
\newblock Effects of an in-car augmented reality system on improving safety of
  younger and older drivers.
\newblock In {\em IEEE International Symposium on Mixed and Augmented Reality
  (ISMAR)}, pp. 59--66, 2013.

\bibitem{5643560}
M.~{Gandy}, R.~{Catrambone}, B.~{MacIntyre}, C.~{Alvarez}, E.~{Eiriksdottir},
  M.~{Hilimire}, B.~{Davidson}, and A.~C. {McLaughlin}.
\newblock Experiences with an {AR} evaluation test bed: Presence, performance,
  and physiological measurement.
\newblock In {\em IEEE International Symposium on Mixed and Augmented Reality
  (ISMAR)}, pp. 127--136, 2010.

\bibitem{3347933}
C.~Gebhardt, B.~Hecox, B.~van Opheusden, D.~Wigdor, J.~Hillis, O.~Hilliges, and
  H.~Benko.
\newblock Learning cooperative personalized policies from gaze data.
\newblock In {\em Proceedings of the ACM Symposium on User Interface Software
  and Technology (UIST)}, pp. 197–--208, 2019.

\bibitem{8613754}
M.~{Geronazzo}, E.~{Sikström}, J.~{Kleimola}, F.~{Avanzini}, A.~{de Götzen},
  and S.~{Serafin}.
\newblock The impact of an accurate vertical localization with {HRTFs} on short
  explorations of immersive virtual reality scenarios.
\newblock In {\em IEEE International Symposium on Mixed and Augmented Reality
  (ISMAR)}, pp. 90--97, 2018.

\bibitem{5643564}
L.~{Gruber}, S.~{Gauglitz}, J.~{Ventura}, S.~{Zollmann}, M.~{Huber},
  M.~{Schlegel}, G.~{Klinker}, D.~{Schmalstieg}, and T.~{Höllerer}.
\newblock The city of sights: Design, construction, and measurement of an
  augmented reality stage set.
\newblock In {\em IEEE International Symposium on Mixed and Augmented Reality
  (ISMAR)}, pp. 157--163, 2010.

\bibitem{7435333}
J.~{Grubert}, T.~{Langlotz}, S.~{Zollmann}, and H.~{Regenbrecht}.
\newblock Towards pervasive augmented reality: Context-awareness in augmented
  reality.
\newblock {\em Transactions on Visualization and Computer Graphics},
  23(6):1706--1724, 2017.

\bibitem{7523400}
K.~{Gupta}, G.~A. {Lee}, and M.~{Billinghurst}.
\newblock Do you see what {I} see? {The} effect of gaze tracking on task space
  remote collaboration.
\newblock {\em Transactions on Visualization and Computer Graphics},
  22(11):2413--2422, 2016.

\bibitem{Harm19a}
C.~Harms and F.~Biocca.
\newblock Internal consistency and reliability of the networked minds measure
  of social presence.
\newblock In M.~Alcaniz and B.~Rey, eds., {\em Proceedings of the International
  Workshop on Presence}, 2004.

\bibitem{Hart88a}
S.~G. Hart and L.~E. Stavenland.
\newblock Development of {NASA-TLX} (task load index): Results of empirical and
  theoretical research.
\newblock In P.~A. Hancock and N.~Meshkati, eds., {\em Human Mental Workload},
  chap.~7, pp. 139--183. Elsevier, North-Holland, 1988.

\bibitem{6671766}
A.~{Hartl}, J.~{Grubert}, D.~{Schmalstieg}, and G.~{Reitmayr}.
\newblock Mobile interactive hologram verification.
\newblock In {\em IEEE International Symposium on Mixed and Augmented Reality
  (ISMAR)}, pp. 75--82, 2013.

\bibitem{5336486}
S.~J. {Henderson} and S.~{Feiner}.
\newblock Evaluating the benefits of augmented reality for task localization in
  maintenance of an armored personnel carrier turret.
\newblock In {\em IEEE International Symposium on Mixed and Augmented Reality
  (ISMAR)}, pp. 135--144, 2009.

\bibitem{8457524}
A.~{Ibrahim}, B.~{Huynh}, J.~{Downey}, T.~{H{\"o}llerer}, D.~{Chun}, and
  J.~{O'Donovan}.
\newblock {ARbis} pictus: A study of vocabulary learning with augmented
  reality.
\newblock {\em Transactions on Visualization and Computer Graphics},
  24(11):2867--2874, 2018.

\bibitem{Ijss08a}
W.~Ijsselsteijn, W.~Van Den~Hoogen, C.~Klimmt, Y.~De~Kort, C.~Lindley,
  K.~Mathiak, K.~Poels, N.~Ravaja, M.~Turpeinen, and P.~Vorderer.
\newblock Measuring the experience of digital game enjoyment.
\newblock In {\em Proceedings of the International Conference on Methods and
  Techniques in Behavioral Research}, pp. 88--89, 2008.

\bibitem{Isen13a}
T.~Isenberg, P.~Isenberg, J.~Chen, M.~Sedlmair, and T.~M{\"o}ller.
\newblock A systematic review on the practice of evaluating visualization.
\newblock {\em Transactions on Visualization and Computer Graphics},
  19(12):2818--2827, 2013.

\bibitem{8613752}
J.~{Jung}, H.~{Lee}, J.~{Choi}, A.~{Nanda}, U.~{Gruenefeld}, T.~{Stratmann},
  and W.~{Heuten}.
\newblock Ensuring safety in augmented reality from trade-off between immersion
  and situation awareness.
\newblock In {\em IEEE International Symposium on Mixed and Augmented Reality
  (ISMAR)}, pp. 70--79, 2018.

\bibitem{Kahn73a}
D.~Kahneman.
\newblock {\em Attention and Effort}.
\newblock Prentice-Hall, Englewood Cliffs, New Jersey, 1973.

\bibitem{Kane07a}
R.~L. Kane, T.~Roebuck-Spencer, P.~Short, M.~Kabat, and J.~Wilken.
\newblock Identifying and monitoring cognitive deficits in clinical populations
  using automated neuropsychological assessment metrics ({ANAM}) tests.
\newblock {\em Archives of Clinical Neuropsychology}, 22:115--126, 2007.

\bibitem{2208641}
A.~Karnik, W.~Mayol-Cuevas, and S.~Subramanian.
\newblock {MUSTARD}: A multi user see through ar display.
\newblock In {\em Proceedings of the ACM Conference on Human Factors in
  Computing Systems (CHI)}, pp. 2541--2550, 2012.

\bibitem{Kenn03a}
R.~S. Kennedy, J.~M. Drexler, D.~E. Compton, K.~M. Stanney, D.~S. Lanham, and
  D.~L. Harm.
\newblock Configural scoring of simulator sickness, cybersickness and space
  adaptation syndrome: Similarities and differences.
\newblock {\em Virtual and adaptive environments: Applications, implications,
  and human performance issues}, p. 247, 2003.

\bibitem{Kim18a}
K.~Kim, M.~Billinghurst, G.~Bruder, H.~B.-L. Duh, and G.~F. Welch.
\newblock Revisiting trends in augmented reality research: A review of the
  2\textsuperscript{nd} decade of {ISMAR} (2008--2017).
\newblock {\em Transactions on Visualization and Computer Graphics},
  24(11):2947--2962, 2018.

\bibitem{8613756}
K.~{Kim}, L.~{Boelling}, S.~{Haesler}, J.~{Bailenson}, G.~{Bruder}, and G.~F.
  {Welch}.
\newblock Does a digital assistant need a body? {The} influence of visual
  embodiment and social behavior on the perception of intelligent virtual
  agents in {AR}.
\newblock In {\em IEEE International Symposium on Mixed and Augmented Reality
  (ISMAR)}, pp. 105--114, 2018.

\bibitem{3300403}
S.~Kim, G.~Lee, W.~Huang, H.~Kim, W.~Woo, and M.~Billinghurst.
\newblock Evaluating the combination of visual communication cues for
  {HMD}-based mixed reality remote collaboration.
\newblock In {\em Proceedings of the ACM Conference on Human Factors in
  Computing Systems (CHI)}, 2019.
\newblock Paper 173.

\bibitem{6948412}
S.~{Kim}, G.~{Lee}, N.~{Sakata}, and M.~{Billinghurst}.
\newblock Improving co-presence with augmented visual communication cues for
  sharing experience through video conference.
\newblock In {\em IEEE International Symposium on Mixed and Augmented Reality
  (ISMAR)}, pp. 83--92, 2014.

\bibitem{Kitc02a}
B.~A. Kitchenham, S.~L. Pfleeger, L.~M. Pickard, P.~W. Jones, D.~C. Hoaglin,
  K.~E. Emam, and J.~Rosenberg.
\newblock Preliminary guidelines for empirical research in software
  engineering.
\newblock {\em Transactions on Software Engineering}, 28(8):721--734, 2002.

\bibitem{5336501}
B.~{Knorlein}, M.~{Di Luca}, and M.~{Harders}.
\newblock Influence of visual and haptic delays on stiffness perception in
  augmented reality.
\newblock In {\em IEEE International Symposium on Mixed and Augmented Reality
  (ISMAR)}, pp. 49--52, 2009.

\bibitem{5336485}
A.~{Kotranza}, D.~{Scott Lind}, C.~M. {Pugh}, and B.~{Lok}.
\newblock Real-time in-situ visual feedback of task performance in mixed
  environments for learning joint psychomotor-cognitive tasks.
\newblock In {\em IEEE International Symposium on Mixed and Augmented Reality
  (ISMAR)}, pp. 125--134, 2009.

\bibitem{6948405}
M.~{Krichenbauer}, G.~{Yamamoto}, T.~{Taketomi}, C.~{Sandor}, and H.~{Kato}.
\newblock Towards augmented reality user interfaces in {3D} media production.
\newblock In {\em IEEE International Symposium on Mixed and Augmented Reality
  (ISMAR)}, pp. 23--28, 2014.

\bibitem{krippendorff2018content}
K.~Krippendorff.
\newblock {\em Content Analysis: An Introduction to its Methodology}.
\newblock Sage Publications, California, 2018.

\bibitem{5643530}
E.~{Kruijff}, J.~E. {Swan}, and S.~{Feiner}.
\newblock Perceptual issues in augmented reality revisited.
\newblock In {\em IEEE International Symposium on Mixed and Augmented Reality
  (ISMAR)}, pp. 3--12, 2010.

\bibitem{Lam12a}
H.~Lam, E.~Bertini, P.~Isenberg, C.~Plaisant, and S.~Carpendale.
\newblock Empirical studies in information visualization: Seven scenarios.
\newblock {\em Transactions on Visualization and Computer Graphics},
  18(9):1520--1536, 2012.

\bibitem{landis1977measurement}
J.~R. Landis and G.~G. Koch.
\newblock The measurement of observer agreement for categorical data.
\newblock {\em Biometrics}, pp. 159--174, 1977.

\bibitem{7523376}
T.~{Langlotz}, M.~{Cook}, and H.~{Regenbrecht}.
\newblock Real-time radiometric compensation for optical see-through
  head-mounted displays.
\newblock {\em Transactions on Visualization and Computer Graphics},
  22(11):2385--2394, 2016.

\bibitem{8613761}
G.~A. {Lee}, T.~{Teo}, S.~{Kim}, and M.~{Billinghurst}.
\newblock A user study on {MR} remote collaboration using live 360 video.
\newblock In {\em IEEE International Symposium on Mixed and Augmented Reality
  (ISMAR)}, pp. 153--164, 2018.

\bibitem{Lewi95a}
J.~R. Lewis.
\newblock {IBM} computer usability satisfaction questionnaires: psychometric
  evaluation and instructions for use.
\newblock {\em International Journal of Human-Computer Interaction},
  7(1):57--78, 1995.

\bibitem{Lind14a}
A.~Lindau, V.~Erbes, S.~Lepa, H.-J. Maempel, F.~Brinkmann, and S.~Weinzierl.
\newblock A spatial audio quality inventory ({SAQI}).
\newblock {\em Acta Acustica united with Acustica}, 100(5):984--994, 2014.

\bibitem{Llob13a}
J.~Llobera, M.~V. Sanchez-Vives, and M.~Slater.
\newblock The relationship between virtual body ownership and temperature
  sensitivity.
\newblock {\em Journal of the Royal Society}, 10(85):1--11, 2013.

\bibitem{Lomb09a}
M.~Lombard, T.~B. Ditton, and L.~Weinstein.
\newblock Measuring presence: the temple presence inventory.
\newblock In {\em Proceedings of the International Workshop on Presence}, pp.
  1--15, 2009.

\bibitem{8613760}
F.~{Lu}, D.~{Yu}, H.~{Liang}, W.~{Chen}, K.~{Papangelis}, and N.~M. {Ali}.
\newblock Evaluating engagement level and analytical support of interactive
  visualizations in virtual reality environments.
\newblock In {\em IEEE International Symposium on Mixed and Augmented Reality
  (ISMAR)}, pp. 143--152, 2018.

\bibitem{6671762}
M.~R. {Marner}, A.~{Irlitti}, and B.~H. {Thomas}.
\newblock Improving procedural task performance with augmented reality
  annotations.
\newblock In {\em IEEE International Symposium on Mixed and Augmented Reality
  (ISMAR)}, pp. 39--48, 2013.

\bibitem{Marr18a}
K.~Marriott, F.~Schreiber, T.~Dwyer, K.~Klein, N.~H. Riche, T.~Itoh,
  W.~Stuerzlinger, and B.~H. Thomas.
\newblock {\em Immersive Analytics}.
\newblock Springer, Cham, 2018.

\bibitem{Mcau89a}
E.~McAuley, T.~Duncan, and V.~V. Tammen.
\newblock Psychometric properties of the intrinsic motivation inventory in a
  competitive sport setting: A confirmatory factor analysis.
\newblock {\em Research Quarterly for Exercise and Sport}, 60(1):48--58, 1989.

\bibitem{mcgrath1995methodology}
J.~E. McGrath.
\newblock Methodology matters: Doing research in the behavioral and social
  sciences.
\newblock In R.~M. Baecker, J.~Grudin, W.~A. Buxton, and S.~Greenberg, eds.,
  {\em Readings in Human--Computer Interaction}, Interactive Technologies, pp.
  152--169. Morgan Kaufmann, Burlington, 1995.

\bibitem{Mckn11a}
D.~H. Mcknight, M.~Carter, J.~B. Thatcher, and P.~F. Clay.
\newblock Trust in a specific technology: An investigation of its components
  and measures.
\newblock {\em Transactions on Management Information Systems},
  2(2):12:1--12:25, 2011.

\bibitem{8466859}
C.~{Merenda}, H.~{Kim}, K.~{Tanous}, J.~L. {Gabbard}, B.~{Feichtl}, T.~{Misu},
  and C.~{Suga}.
\newblock Augmented reality interface design approaches for goal-directed and
  stimulus-driven driving tasks.
\newblock {\em Transactions on Visualization and Computer Graphics},
  24(11):2875--2885, 2018.

\bibitem{Meri18a}
L.~Merino, M.~Ghafari, C.~Anslow, and O.~Nierstrasz.
\newblock A systematic literature review of software visualization evaluation.
\newblock {\em Journal of Systems and Software}, 144:165--180, 2018.

\bibitem{dataset}
L.~Merino, M.~Schwarzl, M.~Kraus, M.~Sedlmair, D.~Schmalstieg, and D.~Weiskopf.
\newblock {Dataset: Evaluating Mixed and Augmented Reality: A Systematic
  Literature Review (2009--2019)}, Mar. 2020. doi: {{%
10\hspace{.1pt}\discretionary{.}{%
}{.}\hspace{.4pt}5281\discretionary{/}{%
}{/}zenodo\hspace{.1pt}\discretionary{.}{%
}{.}\hspace{.4pt}3832114}}


\bibitem{Muld80a}
G.~Mulder.
\newblock {\em The Heart of Mental Effort}.
\newblock PhD thesis, University of Groningen, The Netherlands, 1980.

\bibitem{Munz08a}
T.~Munzner.
\newblock Process and pitfalls in writing information visualization research
  papers.
\newblock In A.~Kerren, J.~T. Stasko, J.-D. Fekete, and C.~North, eds., {\em
  Information Visualization: Human-Centered Issues and Perspectives}, pp.
  134--153. Springer, Berlin, Heidelberg, 2008.

\bibitem{6162874}
T.~{Olsson} and M.~{Salo}.
\newblock Online user survey on current mobile augmented reality applications.
\newblock In {\em IEEE International Symposium on Mixed and Augmented Reality
  (ISMAR)}, pp. 75--84, 2011.

\bibitem{Oppermann:2015:SPA:2702123.2702538}
L.~Oppermann, C.~Putschli, C.~Brosda, O.~Lobunets, and F.~Prioville.
\newblock The smartphone project: An augmented dance performance.
\newblock In {\em Proceedings of the ACM Conference on Human Factors in
  Computing Systems (CHI)}, pp. 2569--2572, 2015.

\bibitem{8115401}
J.~{Orlosky}, P.~{Kim}, K.~{Kiyokawa}, T.~{Mashita}, P.~{Ratsamee},
  Y.~{Uranishi}, and H.~{Takemura}.
\newblock {VisMerge}: Light adaptive vision augmentation via spectral and
  temporal fusion of non-visible light.
\newblock In {\em IEEE International Symposium on Mixed and Augmented Reality
  (ISMAR)}, pp. 22--31, 2017.

\bibitem{7164337}
J.~{Orlosky}, T.~{Toyama}, K.~{Kiyokawa}, and D.~{Sonntag}.
\newblock {ModulAR}: Eye-controlled vision augmentations for head mounted
  displays.
\newblock {\em Transactions on Visualization and Computer Graphics},
  21(11):1259--1268, 2015.

\bibitem{Paas92b}
F.~Paas.
\newblock Training strategies for attaining transfer of problem-solving skill
  in statistics: A cognitive-load approach.
\newblock {\em Journal of Educational Psychology}, 84:429--434, Dec. 1992.

\bibitem{6948411}
T.~{Piumsomboon}, D.~{Altimira}, H.~{Kim}, A.~{Clark}, G.~{Lee}, and
  M.~{Billinghurst}.
\newblock Grasp-shell vs gesture-speech: A comparison of direct and indirect
  natural interaction techniques in augmented reality.
\newblock In {\em IEEE International Symposium on Mixed and Augmented Reality
  (ISMAR)}, pp. 73--82, 2014.

\bibitem{8466636}
T.~{Piumsomboon}, G.~A. {Lee}, B.~{Ens}, B.~H. {Thomas}, and M.~{Billinghurst}.
\newblock Superman vs giant: A study on spatial perception for a multi-scale
  mixed reality flying telepresence interface.
\newblock {\em Transactions on Visualization and Computer Graphics},
  24(11):2974--2982, 2018.

\bibitem{3300458}
T.~Piumsomboon, G.~A. Lee, A.~Irlitti, B.~Ens, B.~H. Thomas, and
  M.~Billinghurst.
\newblock On the shoulder of the giant: A multi-scale mixed reality
  collaboration with 360 video sharing and tangible interaction.
\newblock In {\em Proceedings of the ACM Conference on Human Factors in
  Computing Systems (CHI)}, 2019.
\newblock Paper 228.

\bibitem{8456571}
L.~{Qian}, A.~{Plopski}, N.~{Navab}, and P.~{Kazanzides}.
\newblock Restoring the awareness in the occluded visual field for optical
  see-through head-mounted displays.
\newblock {\em Transactions on Visualization and Computer Graphics},
  24(11):2936--2946, 2018.

\bibitem{6402561}
I.~{Radu} and B.~{MacIntyre}.
\newblock Using children's developmental psychology to guide augmented-reality
  design and usability.
\newblock In {\em IEEE International Symposium on Mixed and Augmented Reality
  (ISMAR)}, pp. 227--236, 2012.

\bibitem{3300774}
I.~Radu and B.~Schneider.
\newblock What can we learn from augmented reality {(AR)}?
\newblock In {\em Proceedings of the ACM Conference on Human Factors in
  Computing Systems (CHI)}, 2019.
\newblock Paper 544.

\bibitem{7523388}
F.~{Rameau}, H.~{Ha}, K.~{Joo}, J.~{Choi}, K.~{Park}, and I.~S. {Kweon}.
\newblock A real-time augmented reality system to see-through cars.
\newblock {\em Transactions on Visualization and Computer Graphics},
  22(11):2395--2404, 2016.

\bibitem{8115408}
H.~{Regenbrecht}, K.~{Meng}, A.~{Reepen}, S.~{Beck}, and T.~{Langlotz}.
\newblock Mixed voxel reality: Presence and embodiment in low fidelity,
  visually coherent, mixed reality environments.
\newblock In {\em IEEE International Symposium on Mixed and Augmented Reality
  (ISMAR)}, pp. 90--99, 2017.

\bibitem{Rege17a}
H.~Regenbrecht, T.~Schubert, C.~Botella, and R.~Ba{\~n}os.
\newblock Mixed reality experience questionnaire ({MREQ})-reference.
\newblock Technical report, University of Otago, 2017.

\bibitem{8462799}
C.~{Reichherzer}, A.~{Cunningham}, J.~{Walsh}, M.~{Kohler}, M.~{Billinghurst},
  and B.~H. {Thomas}.
\newblock Narrative and spatial memory for jury viewings in a reconstructed
  virtual environment.
\newblock {\em Transactions on Visualization and Computer Graphics},
  24(11):2917--2926, 2018.

\bibitem{Schmalstieg2016}
D.~Schmalstieg and T.~H\"ollerer.
\newblock {\em Augmented Reality: Principles and Practice}.
\newblock Addison-Wesley Professional, Boston, 2016.

\bibitem{Schu01a}
T.~Schubert, F.~Friedmann, and H.~Regenbrecht.
\newblock The experience of presence: Factor analytic insights.
\newblock {\em Presence: Teleoperators and Virtual Environments}, 10:266--281,
  June 2001.

\bibitem{5336484}
B.~{Schwerdtfeger}, R.~{Reif}, W.~A. {Gunthner}, G.~{Klinker}, D.~{Hamacher},
  L.~{Schega}, I.~{Bockelmann}, F.~{Doil}, and J.~{Tumler}.
\newblock Pick-by-vision: A first stress test.
\newblock In {\em IEEE International Symposium on Mixed and Augmented Reality
  (ISMAR)}, pp. 115--124, 2009.

\bibitem{sharp2008vak}
J.~G. Sharp, R.~Bowker, and J.~Byrne.
\newblock {VAK} or {VAK}-uous? {Towards} the trivialisation of learning and the
  death of scholarship.
\newblock {\em Research Papers in Education}, 23(3):293--314, 2008.

\bibitem{8943618}
X.~{Shi}, J.~{Pan}, Z.~{Hu}, J.~{Lin}, S.~{Guo}, M.~{Liao}, Y.~{Pan}, and
  L.~{Liu}.
\newblock Accurate and fast classification of foot gestures for virtual
  locomotion.
\newblock In {\em IEEE International Symposium on Mixed and Augmented Reality
  (ISMAR)}, pp. 178--189, 2019.

\bibitem{Slat94a}
M.~Slater, M.~Usoh, and A.~Steed.
\newblock Depth of presence in virtual environments.
\newblock {\em Presence: Teleoperators and Virtual Environments},
  3(2):130--144, 1994.

\bibitem{6948430}
W.~{Steptoe}, S.~{Julier}, and A.~{Steed}.
\newblock Presence and discernability in conventional and non-photorealistic
  immersive augmented reality.
\newblock In {\em IEEE International Symposium on Mixed and Augmented Reality
  (ISMAR)}, pp. 213--218, 2014.

\bibitem{swan2005survey}
J.~E. Swan and J.~L. Gabbard.
\newblock Survey of user-based experimentation in augmented reality.
\newblock In {\em Proceedings of International Conference on Virtual Reality},
  vol.~22, pp. 1--9, 2005.

\bibitem{3300431}
T.~Teo, L.~Lawrence, G.~A. Lee, M.~Billinghurst, and M.~Adcock.
\newblock Mixed reality remote collaboration combining 360 video and {3D}
  reconstruction.
\newblock In {\em Proceedings of the ACM Conference on Human Factors in
  Computing Systems (CHI)}, 2019.
\newblock Paper 201.

\bibitem{8943651}
S.~{Thompson}, A.~{Chalmers}, and T.~{Rhee}.
\newblock Real-time mixed reality rendering for underwater 360° videos.
\newblock In {\em IEEE International Symposium on Mixed and Augmented Reality
  (ISMAR)}, pp. 74--82, 2019.

\bibitem{Tomi10a}
T.~{Tominaga}, T.~{Hayashi}, J.~{Okamoto}, and A.~{Takahashi}.
\newblock Performance comparisons of subjective quality assessment methods for
  mobile video.
\newblock In {\em Proceedings of the International Workshop on Quality of
  Multimedia Experience}, pp. 82--87, 2010.

\bibitem{Boxt93a}
A.~Van~Boxtel and M.~Jessurun.
\newblock Amplitude and bilateral coherency of facial and jaw-elevator {EMG}
  activity as an index of effort during a two-choice serial reaction task.
\newblock {\em Psychophysiology}, 30(6):589--604, 1993.

\bibitem{8493594}
B.~{Volmer}, J.~{Baumeister}, S.~{Von Itzstein}, I.~{Bornkessel-Schlesewsky},
  M.~{Schlesewsky}, M.~{Billinghurst}, and B.~H. {Thomas}.
\newblock A comparison of predictive spatial augmented reality cues for
  procedural tasks.
\newblock {\em Transactions on Visualization and Computer Graphics},
  24(11):2846--2856, 2018.

\bibitem{Vord04a}
P.~Vorderer, W.~Wirth, F.~Gouveia, F.~Biocca, T.~Saari, L.~J{\"a}ncke,
  S.~B{\"o}cking, H.~Schramm, A.~Gysbers, T.~Hartmann, C.~Klimmt, J.~Laarni,
  N.~Ravaja, A.~Sacau, T.~Baumgartner, and P.~J{\"o}ncke.
\newblock {MEC} spatial presence questionnaire ({MEC-SPQ}): Short documentation
  and instructions for application.
\newblock {\em Report to the European Community, Project Presence: MEC
  (IST-2001-37661)}, 2004.

\bibitem{Wats88a}
D.~Watson, L.~Anna~Clark, and A.~Tellegen.
\newblock Development and validation of brief measures of positive and negative
  affect: The {PANAS} scales.
\newblock {\em Journal of Personality and Social Psychology}, 54:1063--1070,
  June 1988.

\bibitem{8115412}
C.~A. {Wiesner}, M.~{Ruf}, D.~{Sirim}, and G.~{Klinker}.
\newblock {3D-FRC}: Depiction of the future road course in the head-up-display.
\newblock In {\em IEEE International Symposium on Mixed and Augmented Reality
  (ISMAR)}, pp. 136--143, 2017.

\bibitem{7523389}
S.~{Willi} and A.~{Grundhöfer}.
\newblock Spatio-temporal point path analysis and optimization of a
  galvanoscopic scanning laser projector.
\newblock {\em Transactions on Visualization and Computer Graphics},
  22(11):2377--2384, 2016.

\bibitem{8943645}
Y.~{Wu}, L.~{Chan}, and W.~{Lin}.
\newblock Tangible and visible {3D} object reconstruction in augmented reality.
\newblock In {\em IEEE International Symposium on Mixed and Augmented Reality
  (ISMAR)}, pp. 26--36, 2019.

\bibitem{3300674}
W.~Xu, H.-N. Liang, Y.~Zhao, D.~Yu, and D.~Monteiro.
\newblock {DMove}: Directional motion-based interaction for augmented reality
  head-mounted displays.
\newblock In {\em Proceedings of the ACM Conference on Human Factors in
  Computing Systems (CHI)}, 2019.
\newblock Paper 444.

\bibitem{Yesh08a}
Y.~Yeshurun, M.~Carrasco, and L.~T. Maloney.
\newblock Bias and sensitivity in two-interval forced choice procedures: Tests
  of the difference model.
\newblock {\em Vision Research}, 48(17):1837--1851, 2008.

\bibitem{8456570}
D.~{Yu}, K.~{Fan}, H.~{Zhang}, D.~{Monteiro}, W.~{Xu}, and H.~{Liang}.
\newblock {PizzaText}: Text entry for virtual reality systems using dual
  thumbsticks.
\newblock {\em Transactions on Visualization and Computer Graphics},
  24(11):2927--2935, 2018.

\bibitem{7781773}
J.~{Zhang}, A.~{Ogan}, T.~{Liu}, Y.~{Sung}, and K.~{Chang}.
\newblock The influence of using augmented reality on textbook support for
  learners of different learning styles.
\newblock In {\em IEEE International Symposium on Mixed and Augmented Reality
  (ISMAR)}, pp. 107--114, 2016.

\bibitem{Zhou08a}
F.~Zhou, H.~B.-L. Duh, and M.~Billinghurst.
\newblock Trends in augmented reality tracking, interaction and display: A
  review of ten years of {ISMAR}.
\newblock In {\em IEEE International Symposium on Mixed and Augmented Reality
  (ISMAR)}, pp. 193--202, 2008.

\bibitem{Zijl93a}
F.~Zijlstra.
\newblock {\em Efficiency in Work Behavior: A Design Approach for Modern
  Tools}.
\newblock PhD thesis, Technical University of Delft, 1993.

\end{thebibliography}
\end{document}